\documentclass[a4paper,11pt]{article}
\pdfoutput=1 

\newcommand{\degC}{$^{\circ}$C~}

\newcommand{\neut}{$\rm n_{eq}/cm^2$}
\newcommand{\vendorH}{HPK}
\newcommand{\vendorF}{FBK}

\usepackage{orcidlink}
\usepackage{lineno}
\usepackage{setspace}

\usepackage{standalone}
\usepackage{siunitx}
\usepackage{import}
\usepackage{tikz}
\usepackage{graphicx}
\usepackage{subcaption}

\usepackage[version=4]{mhchem}

\usepackage{multirow}

\usepackage{jinstpub} 
\usepackage{cleveref}
\title{Integration of thermo-electric coolers into the CMS MTD SiPM arrays for operation under high neutron fluence}

\newcommand\dnote[2][]{%
\setcounter{footnote}{1}
\renewcommand{\thefootnote}{\fnsymbol{footnote}}
\if!#1!%
\stepcounter{footnote}\footnotetext{#2}%
\else%
{\renewcommand\thefootnote{#1}%
\footnotetext{#2}}%
\fi}

\author[a]{A.~Bornheim,}

\author[b]{W.~Lustermann,}
\author[b]{K.~Stachon,}

\author[c]{G.~Reales~Gutiérrez,}

\author[d]{A.~Benaglia,}
\author[d, h]{F.~De~Guio,}
\author[d, h]{A.~Ghezzi,}
\author[d,h,{\dagger}\dnote{Corresponding authors.}]{M.~T.~Lucchini,}
\author[d]{M.~Malberti,}
\author[d,h]{S.~Palluotto,}
\author[d,h]{T.~Tabarelli~de~Fatis,}

\author[e]{M.~Benettoni,}
\author[e,i]{R.~Carlin,}
\author[e,i]{M.~Tosi,}
\author[e,i]{R.~Rossin,}

\author[f]{P.~Meridiani,}
\author[f,l]{R.~Paramatti,}
\author[f,l]{F.~Santanastasio,}

\author[g]{J.~C.~Silva,}
\author[g]{J.~Varela,}

\author[j,{\dagger}]{A.~Heering,}
\author[j,1]{A.~Karneyeu\orcidlink{0000-0001-9983-1004},}
\author[j,1]{Y.~Musienko\orcidlink{0009-0006-3545-1938},}
\author[j]{M.~Wayne,}

\author[k]{T.~Anderson,}
\author[k]{B.~Cox,}
\author[k]{C.E.~Perez~Lara,}
\author[k]{A.~Ledovskoy,}
\author[k]{S.~White,}

\author[m]{I.~Schmidt}

\affiliation[a]{California Institute of Technology, \\ 1200 E California Blvd, Pasadena, USA}

\affiliation[b]{ETH Zurich - Institute for Particle Physics and Astrophysics (IPA), \\   Otto-Stern-Weg 5, Zurich, Switzerland}

\affiliation[c]{Delft University of Technology,\\ Mekelweg 5, 2628CD, Delft, The Netherlands,}%

\affiliation[d]{INFN Sezione di Milano-Bicocca, \\   Piazza della Scienza 3, Milano, Italy}

\affiliation[e]{INFN Sezione di Padova, \\ Via Marzolo 8, Padova, Italy}

\affiliation[f]{INFN Sezione di Roma, \\ Piazzale Aldo Moro 2, Rome, Italy}

\affiliation[g]{Laboratório de Instrumentação e Física Experimental de Partículas, \\ Av. Elias Garcia 14, Lisboa, Portugal}

\affiliation[h]{Università di Milano-Bicocca, \\   Piazza della Scienza 3, Milano, Italy}

\affiliation[i]{Università di Padova, \\ Via Marzolo 8, Padova, Italy}

\affiliation[j]{University of Notre Dame, \\Notre Dame, USA}

\affiliation[k]{University of Virginia, \\  382 McCormick Rd, Charlottesville, USA}

\affiliation[l]{Sapienza Università di Roma, \\ Piazzale Aldo Moro 2, Rome, Italy}

\affiliation[m]{The University of Iowa, \\Iowa City, USA}

\affiliation[1]{Also at an institute or an international laboratory covered by a cooperation agreement with CERN}

\emailAdd{Marco.Lucchini@cern.ch}
\emailAdd{Arjan.Heering@cern.ch}

\abstract{
The barrel section of the novel MIP Timing Detector (MTD) will be constructed as part of the upgrade of the CMS experiment to provide a time resolution for single charged tracks in the range of $30-60$~ps using LYSO:Ce crystal arrays read out with Silicon Photomultipliers (SiPMs).
A major challenge for the operation of such a detector is the extremely high radiation level, of about $2\times10^{14}$~1~MeV(Si)~Eqv.~n/cm$^2$, that will be integrated over a decade of operation of the High Luminosity Large Hadron Collider (HL-LHC).
Silicon Photomultipliers exposed to this level of radiation have shown a strong increase in dark count rate and radiation damage effects that also impact their gain and photon detection efficiency. For this reason during operations the whole detector is cooled down to about $-35$\degC.
In this paper we illustrate an innovative and cost-effective solution to mitigate the impact of radiation damage on the timing performance of the detector, by integrating small thermo-electric coolers (TECs) on the back of the SiPM package.
This additional feature, fully integrated as part of the SiPM array, enables a further decrease in operating temperature down to about $-45^{\circ}$C. This leads to a reduction by a factor of about two in the dark count rate without requiring additional power budget, since the power required by the TEC is almost entirely offset by a decrease in the power required for the SiPM operation due to leakage current.
In addition, the operation of the TECs with reversed polarity during technical stops of the accelerator can raise the temperature of the SiPMs up to 60\degC (about 50\degC higher than the rest of the detector), thus accelerating the annealing of radiation damage effects and partly recovering the SiPM performance.
}

\keywords{SiPMs, Thermo-Electric Coolers, Radiation Damage, Cooling, Timing Detectors}

\begin{document}

\maketitle

\flushbottom
\section{Introduction}

The CMS detector is planning for a major upgrade during the long shutdown (LS3) which will start around 2026. The upgrade will allow the detector to maintain its event reconstruction capabilities during the High Luminosity phase of the Large Hadron Collider (HL-LHC) addressing the major challenges of about a factor ten times higher integrated luminosity and radiation levels \cite{Butler:2020886}.
The MIP Timing Detector (MTD) is a novel sub-detector, which will be inserted between the silicon tracker and the electromagnetic calorimeter, designed to tag the time of arrival of charged particles with a time resolution in the range of $30-60$~ps \cite{CMS_MTD_TDR}. This additional capability will provide powerful information to help disentangle the overlapping trajectories of charged tracks and mitigate the effects of pile-up on the reconstruction of physics events. This additional time-of-flight information will also enable new searches for long-lived particles and extend the domain of particle identification in the Heavy Ion program.

\begin{figure}[!htbp]
    \centering
    \includegraphics[width=0.99\linewidth]{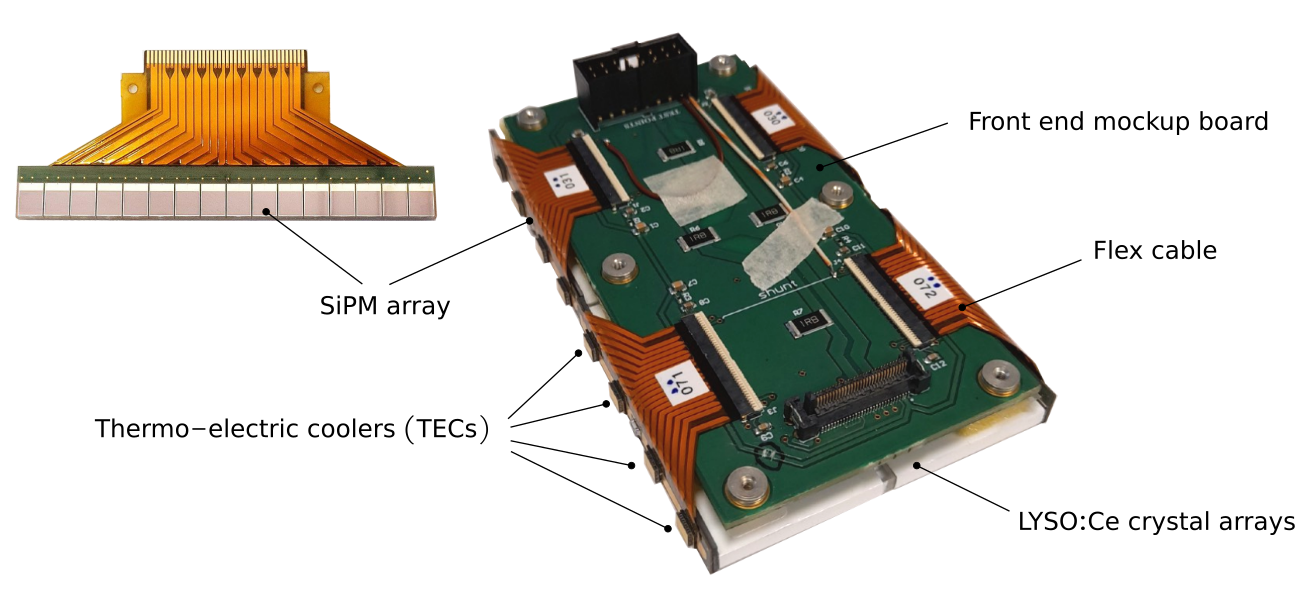}
    \caption{Pictures of SiPM array (front side) and of a BTL detector module, before insertion inside a copper housing, consisting of: 2 LYSO:Ce crystal arrays, 4 SiPM arrays including flex cable for signal routing to the front-end board, 4 thermo-electric coolers (TECs) and one RTD temperature sensor on each array.}
    \label{fig:btl_module}
\end{figure}

\begin{figure}[!htbp]
    \centering
    \includegraphics[width=0.99\linewidth]{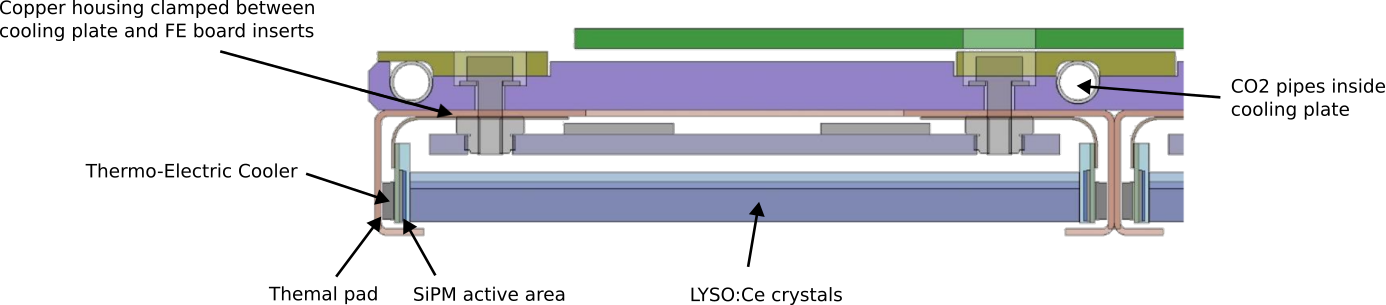}
    \caption{Transverse section of a BTL detector module attached to the cooling plate.}
    \label{fig:btl_module_section}
\end{figure}

The barrel section of the MTD, the Barrel Timing Layer (BTL), is based on LYSO:Ce crystals \cite{Addesa_2022} arranged in arrays of 16 units, readout by packaged arrays of Silicon Photomultipliers (SiPM). The signals from the SiPMs are routed through a custom flex cable to the front-end electronics where a dedicated ASIC (TOFHIR) performs a measurement of time with a fixed threshold discriminator and of the integrated charge as described in \cite{TOFHIR2_Proceeding}.
Two arrays, connected to a front-end electronics board (as shown in figure~\ref{fig:btl_module}), are inserted into a copper housing, to form a detector module. Modules are then connected to an aluminum plate, which during operations is kept at a temperature of $-35^{\circ}$C by a CO$_2$ evaporative cooling system as shown in figure~\ref{fig:btl_module_section} and described in the MTD Technical Design Report \cite{CMS_MTD_TDR}.

Crystals and SiPMs will be exposed throughout the entire operation of the HL-LHC to accumulated radiation levels of 50~kGy of ionizing dose and a neutron fluence of $2\times 10^{14}$~\neut. The notation \neut ~is used here and in the following, as abbreviation of 1~MeV(Si)~Eqv.~n/cm$^2$, to indicate the non-ionizing energy loss (NIEL) damage equivalent to a given flux of 1~MeV neutrons in silicon, which can produce bulk damage by displacing atoms out of their lattice site generating crystal defects.
To our knowledge no other large area experiment has ever used SiPMs in such a harsh radiation environment.
The effects of radiation damage on the SiPMs after these radiation levels have been studied in detail on devices produced by two manufacturers: Hamamatsu Photonics (\vendorH) and Fondazione Bruno Kessler (\vendorF). As described in \cite{GARUTTI201969, Heering2016111}, the first order effect is an increase of the dark count rate (DCR), linearly with the neutron fluence up to about $5\times 10^{13}$~\neut. Beyond this fluence, additional effects start to become sizable. For instance, after a fluence of $2\times 10^{14}$~\neut~an increase of the breakdown voltage by 0.6 and 2.0~V for \vendorF/\vendorH, respectively, occurs and a reduction of the signal amplitude of about 25\% around 1~V over-voltage was measured, suggesting a possible reduction of the SiPM gain and photon detection efficiency (PDE).
A change of the dark count rate dependence on over-voltage was also observed, suggesting that a modification to the silicon doping structure could be happening \cite{Musienko_2020} or that the after-pulsing probability has increased.
After such fluences, a dark count rate in the range 10-100 GHz can be reached depending on the particular operating temperature, over-voltage and annealing history. All these effects have a direct impact on the time resolution of the BTL detector since they introduce additional noise (time jitter) and reduce the amplitude of the signal as described in \cite{CMS_MTD_TDR}. Measurements presented in this paper were obtained with SiPMs having a cell size of $15~\rm \mu m$ but similar results and arguments are valid for SiPMs of larger cell sizes (e.g. $20, 25, 30~\rm \mu m$).

Another major challenge of operating SiPMs under these conditions is the very large power dissipation (up to 50~mW for a 9~mm$^2$ SiPM) which, if not properly accounted for, can lead to undesired self-heating effects with impact on the operating over-voltage \cite{LUCCHINI2020164300}. In particular, an increase of the SiPM temperature by 1\degC corresponds to a few tens of mV increase of the breakdown voltage ($V_{br}$). If the bias voltage ($V_{bias}$) is not adjusted accordingly, the effective over-voltage ($OV = V_{bias} - V_{br}$) is reduced with a non negligible impact on the gain and PDE especially at low operating over-voltages (around 1~V). For this reason a custom substrate for the SiPM arrays was designed by the manufacturers to provide a small thermal resistivity ($2.5-4$~W/K) and thus uniformity of the temperature along the array better than 0.5\degC (a pure ceramic substrate is used by \vendorF~ and a mixed PCB+AlN substrate by \vendorH). 
In the original BTL design \cite{CMS_MTD_TDR} the SiPM package was directly in contact with the copper housing. In the current design, small thermo-electric coolers (TECs) are re-flow soldered on the back of the SiPM package. The TECs behave effectively as an active thermal interface between the SiPM package and copper housing, and as we will show in this paper, give the opportunity to efficiently manage the heat generated by SiPM. A room temperature sensor (RTD) is also included on the rear side of the SiPM array to monitor the temperature more precisely.


The integration of the TECs provide two key opportunities to tackle the high level of dark count rate due to irradiation: the SiPM temperature can be further lowered compared to the temperature of the detector cooling plate (i.e. down to $-45$\degC, reducing the dark count rate by almost a factor of two) during operation, while the SiPM temperature can also be increased up to about 60\degC when the MTD detector is not operational (e.g. during yearly shutdowns of the LHC accelerator and other technical stops) and the cold plate is at 10\degC.
Since the generation rate of dark counts in the silicon drops with temperature and the radiation induced defects introduced in the silicon lattice are thermally annealed at a faster rate for higher temperatures, both the cooling and heating capabilities of the TECs can be exploited to mitigate the SiPM dark count rate as discussed in section~\ref{sec:sipm_raddam}. In sections~\ref{sec:integration}, \ref{sec:model_opt} and \ref{sec:tec_in_btl} we illustrate the technological choices and optimization that led to a successful implementation of this technology on the SiPM arrays for the BTL detector as well as the thermal performance results achieved. The results of dedicated longevity, reliability and radiation tolerance tests performed on the TECs are discussed in section~\ref{sec:reliability}.

\section{Radiation damage and thermal annealing effects on SiPMs}\label{sec:sipm_raddam}
Effects of radiation damage after exposure to neutron fluences up to $3\times10^{14}$~\neut~have been studied on a set of SiPM arrays from different manufacturers. The potential of exploiting thermal annealing to reduce the radiation induced dark count rate has also been studied systematically by keeping the SiPMs in an oven at temperatures up to 70\degC for long periods of time.

\begin{figure}[!tbp]
    \centering
    \includegraphics[width=0.495\linewidth]{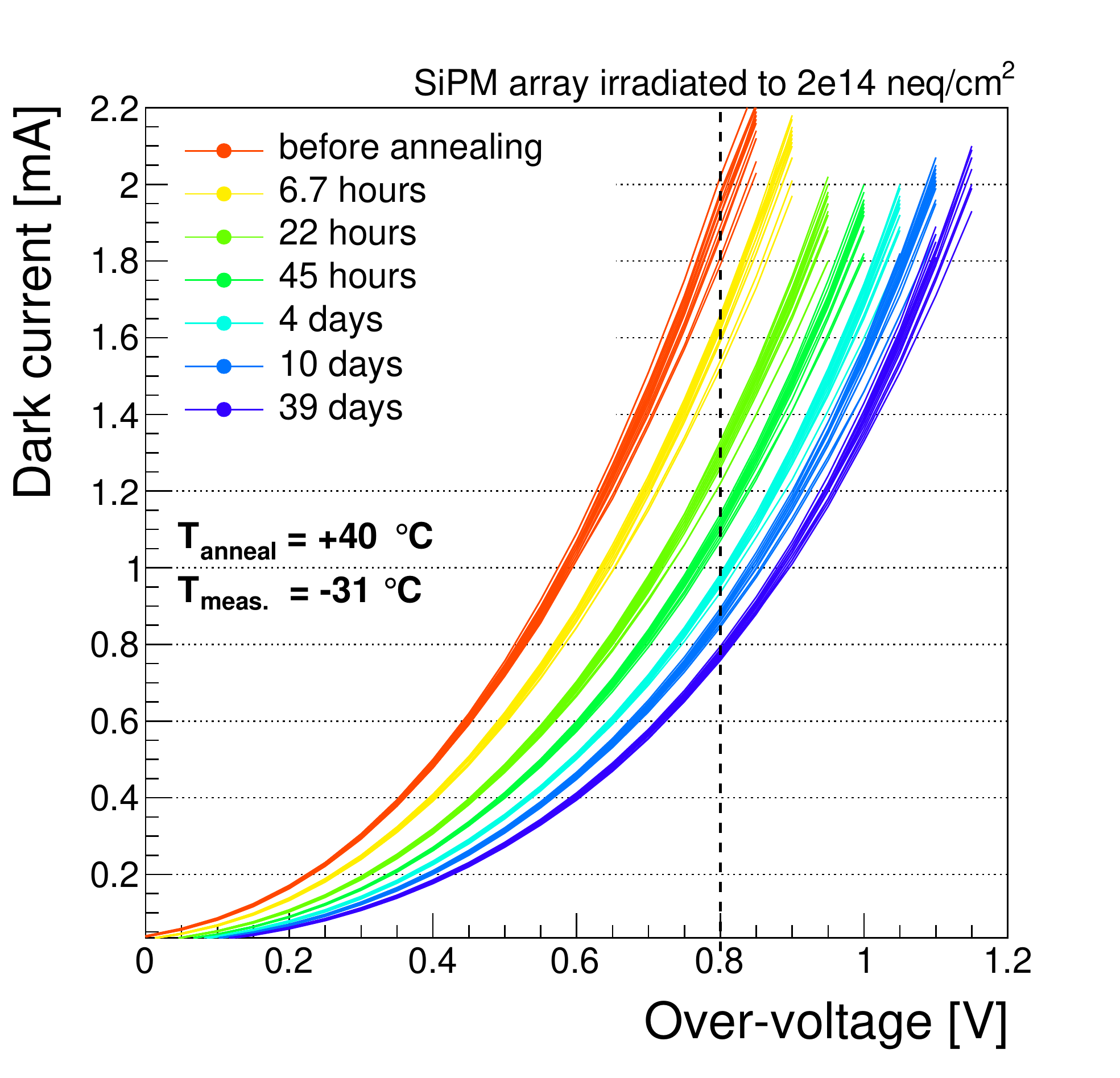}  
    \includegraphics[width=0.495\linewidth]{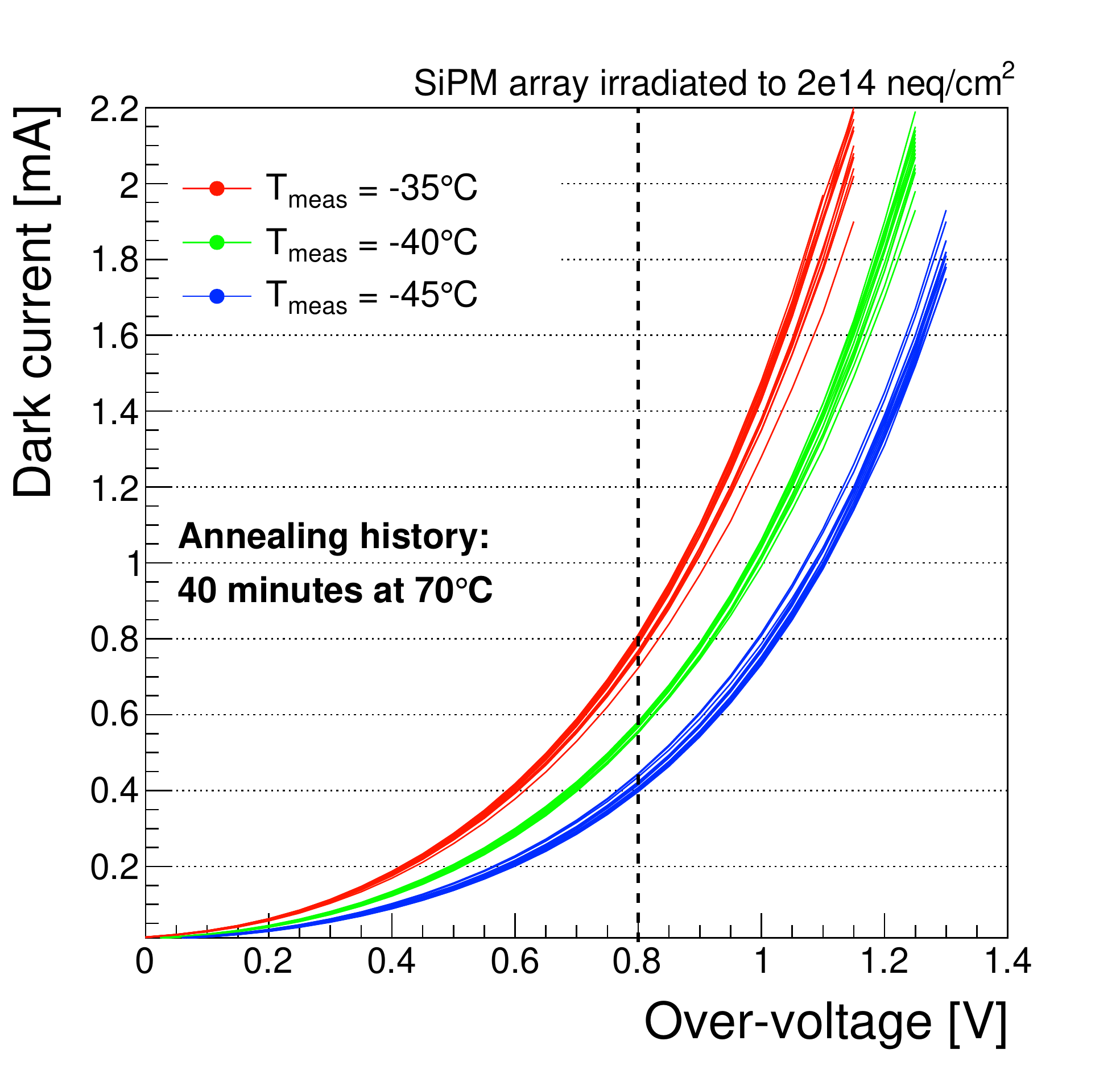}
    \caption{Left: dark current as a function of over-voltage measured at $-31$\degC across all 16 channels of a SiPM array irradiated to $2\times10^{14}$~\neut~after different periods of annealing at 40\degC. Right: dark current of all 16 channels of a SiPM array irradiated to $2\times10^{14}$~\neut~ and annealed for 40 minutes at 70\degC, measured at different SiPM temperatures from $-35$\degC to $-45$\degC. The vertical dashed line at 0.8V over-voltage serves the purpose to guide the eye to assess the impact of dark current reduction from annealing and lower operating temperature.}
    \label{fig:sipm_iv_curves}
\end{figure}

Figure~\ref{fig:sipm_iv_curves} shows the dark current as a function of over-voltage for all the 16 channels of a pair of SiPM arrays irradiated to $2\times10^{14}$~\neut~at the JSI irradiation facility in Ljubljana. One array was placed in an oven at a controlled temperature of 40\degC for various intervals of time, up to a total of about 40 days, and the dark current was measured at $-31$\degC after each annealing step. All SiPM channels within the array show a consistent decrease of the dark current. The right plot of figure~\ref{fig:sipm_iv_curves} illustrates instead the impact of a lower SiPM operating temperature on the dark current measured on a SiPM array which has been briefly annealed after irradiation for 40 minutes at 70\degC and then measured at an operating temperature of $-35$, $-40$ and $-45$\degC.

\begin{figure}[!tbp]
    \centering
    \includegraphics[width=0.495\linewidth]{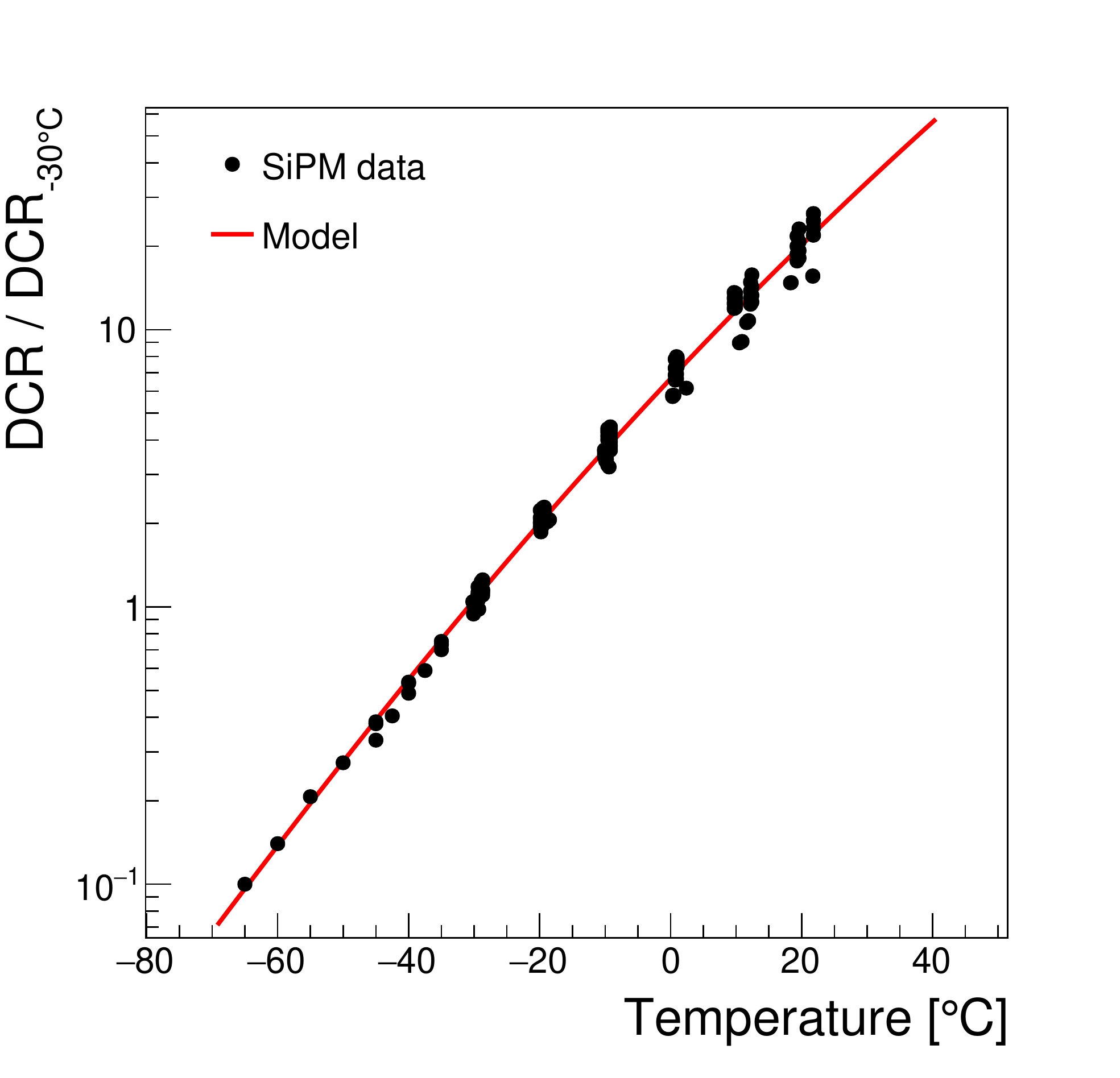}  
    \includegraphics[width=0.495\linewidth]{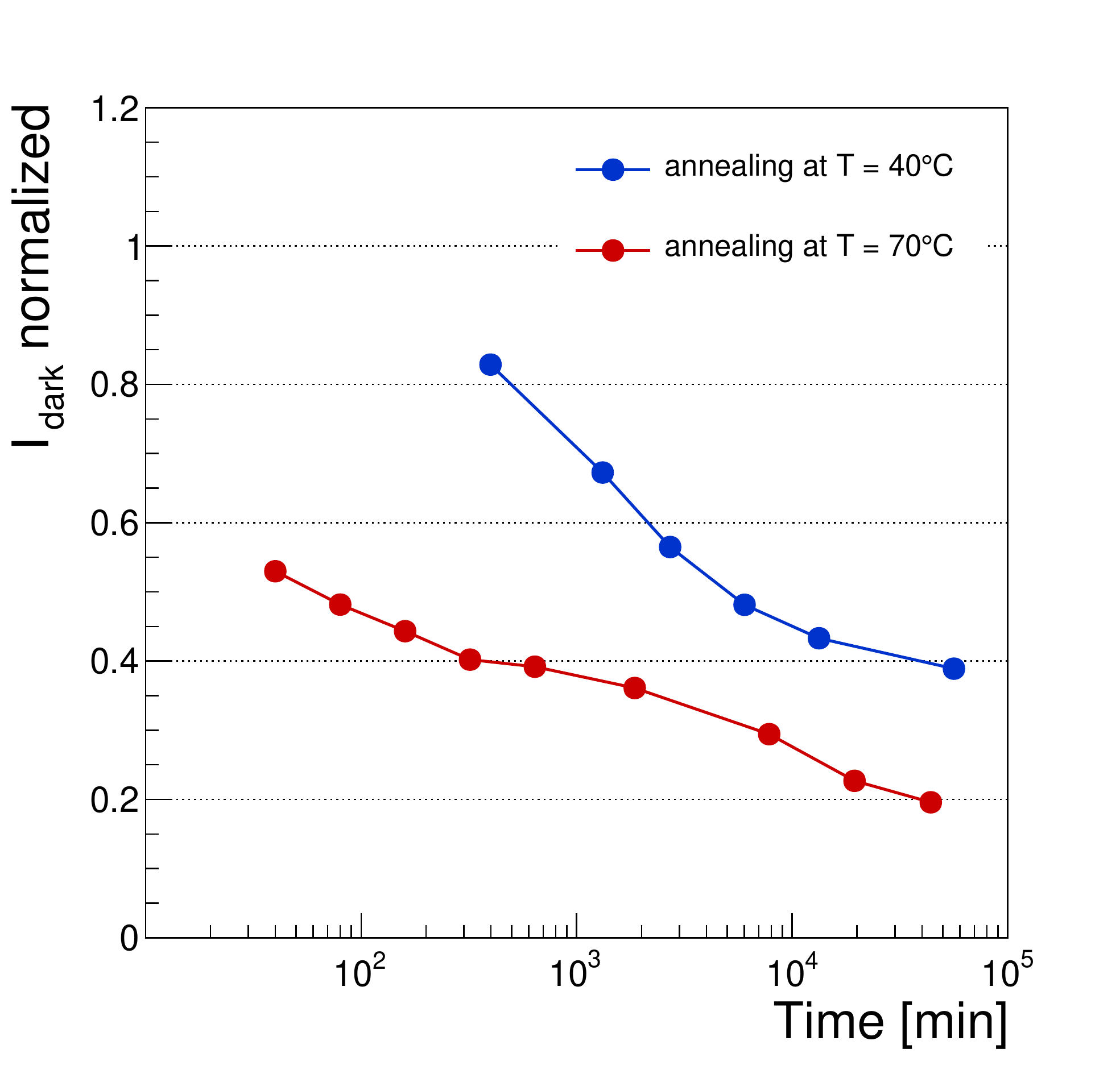}
    \caption{Left: dark count rate evaluated at 1~V over-voltage at a certain temperature T, relative to the dark count rate at $T=-30^{\circ}$C. Data points include results from a set of SiPMs irradiated to various fluences ($1\times10^{13}, 5\times10^{13}, 2\times10^{14}$ and $3\times10^{14}$~\neut) interpolated with the parameterization described in Eq.~\ref{eq:dcr_vs_temp}. Right: reduction of dark current over time due to thermal annealing at 40\degC and 70\degC normalized to the dark current measured after irradiation.}
    \label{fig:sipm_dcr_annealing}
\end{figure}

Figure~\ref{fig:sipm_dcr_annealing} summarizes the results of similar tests performed on a larger set of irradiated SiPMs and over a wider range of operating temperatures ($-60/+30$\degC) and at different annealing temperatures (40 and 70\degC). 
As shown in the left plot of figure~\ref{fig:sipm_dcr_annealing}, the dark count rate evaluated at 1~V over-voltage is reduced by a factor of about 1.9 every 10\degC \cite{HEERING2019671} regardless of the amount of radiation received since the data points include a set of SiPMs irradiated to different fluences ($1\times10^{13}$, $5\times10^{13}$, $2\times 10^{14}$ and $3\times 10^{14}$~\neut) according to the following equation \cite{GARUTTI201969}:

\begin{equation}\label{eq:dcr_vs_temp}
  \rm  \frac{\rm DCR}{\rm DCR (-30^{\circ}C) } = N (1 + \Gamma) T^2 \exp\left(-\frac{E_g/2 + \Delta}{kT}\right)
\end{equation}
where $\rm{E_g}$ is the band-gap for silicon ($\rm E_g(T) = 1.1692 - 4.9 \times 10^{-4}~T^2/(T + 655)$) expressed in eV and $\Gamma$ accounts for effect of tunnelling and depends on the effective electric field strength $\rm{F_{eff}}$, as 

\begin{equation}
\rm
\Gamma = \frac{\gamma \rm{F_{eff}}}{(kT)^{3/2}}\exp\left[\left(\frac{\gamma \rm{F_{eff}}}{(kT)^{3/2}}\right)^2\right].
\end{equation}
The other parameters have been obtained from the fit to the experimental data and correspond to $\Delta = -0.1505 \pm 0.0009$ [eV] and $\rm \gamma F_\mathrm{eff} = (45.0 \pm  0.6)\times 10^{-4}$~[eV$^{3/2}$], where $\gamma$ is a constant~\cite{Hurkx}. The value for the coefficient $\rm N=572.6$~[K$^{-2}$] is obtained from the requirement for the ratio to be equal to 1 at $\rm T=-30$\degC.
%
%
It should be noted that the dependence of dark counts on temperature can vary depending on the specific SiPM technology of a certain producer. We observed a coefficient of about 1.9 every 10\degC for \vendorH~ and about 1.85 every 10\degC for \vendorF~. A slight increase of the temperature coefficient up to about 2.0 was also observed after thermal annealing of the SiPMs.

The annealing behavior in SiPMs was found to follow the same dependence on time and temperature as that of other silicon devices which was broadly tested and documented in \cite{MollThesis}. For instance, the dark current is reduced to about 0.75 of its initial value after 16 hours at 40\degC while it is reduced to 0.38 after the same time for a temperature of 70\degC (thus about a factor of 2 more), as shown in figure~\ref{fig:sipm_dcr_annealing}. The initial value used for the current normalization corresponds to the dark current before any annealing, measured as soon as the SiPMs were received back from the irradiation facility.

\begin{figure}[!tbp]
    \centering
    \includegraphics[width=0.8\linewidth]{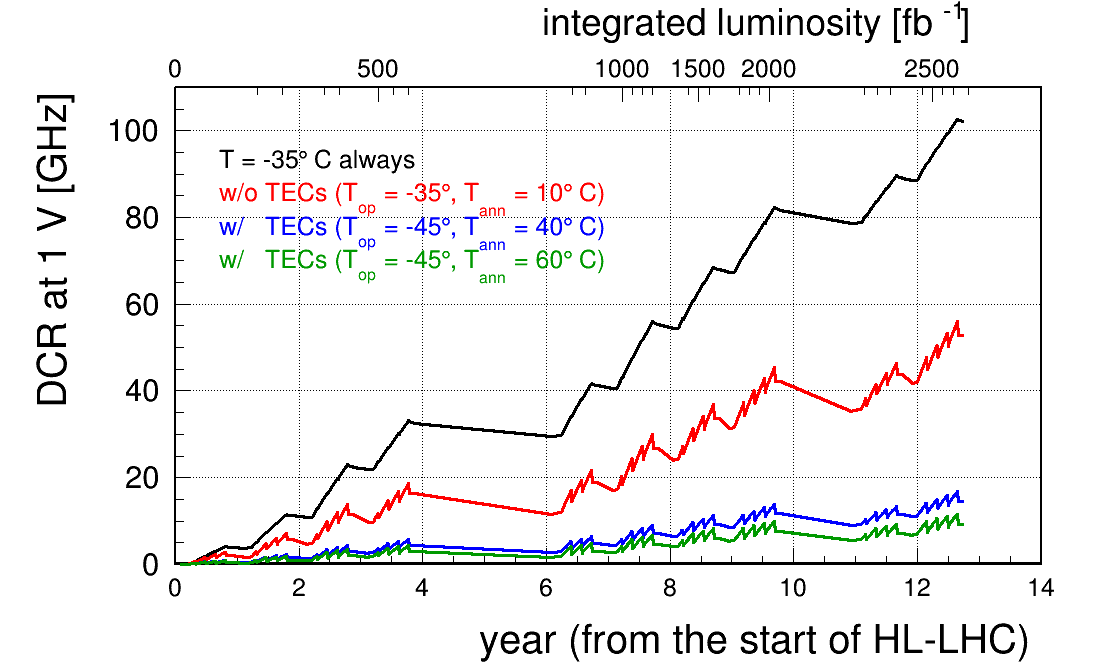}  
    \caption{Evolution of dark count rate at fixed operating over-voltage of 1~V throughout the detector operation assuming the current baseline running plan of the HL-LHC \cite{lhc_schedule} and for different scenarios of SiPM temperature during operation (T$\rm{_{op}}$) and during technical stops (T$\rm{_{ann}}$) when annealing can be accelerated.}
    \label{fig:dcr_annealing}
\end{figure}

During the HL-LHC technical stops, the temperature of the BTL cold plate can be increased from $-35$\degC to a maximum of about $+10$\degC compatibly with the capabilities of the CO$_2$ cooling system and with an acceptable environmental heat loss to the tracker volume.
The biasing of the TECs (if operated in reverse voltage compared to normal operation), however, allows an additional positive temperature gradient between the SiPM and the copper housing of about 50\degC such that a temperature of the SiPM of about 60\degC can be reached. 
This way, high temperature annealing can be performed exploiting the HL-LHC technical stops and long shutdowns, with beneficial impact on the dark count rate and the timing performance.
The expected dark count rate for one SiPM operated at 1~V over-voltage is shown in figure~\ref{fig:dcr_annealing}, throughout the entire detector operation and for a few different scenarios of the operating and annealing temperature. The decrease of operating temperature has obviously a large impact but also does the possibility to increase the detector temperature during technical stops from $-35$\degC to 10, 40 or 60 \degC. For instance, annealing at 60\degC yields approximately a 50\% lower dark count rate compared to annealing at 40\degC for a maximum of 10~GHz instead of 15~GHz at the end of detector operation.

The inclusion of small thermo-electric coolers on the back of the SiPM package thus represents an excellent tool to exploit both the radiation damage annealing and dark count reduction at lower temperature to enable SiPM operation at the unprecedented radiation levels of the MTD.

\section{Integration of TECs into SiPM arrays and proof-of-principle}\label{sec:integration}

\begin{figure}[!tbp]
    \centering
    \includegraphics[width=0.99\linewidth]{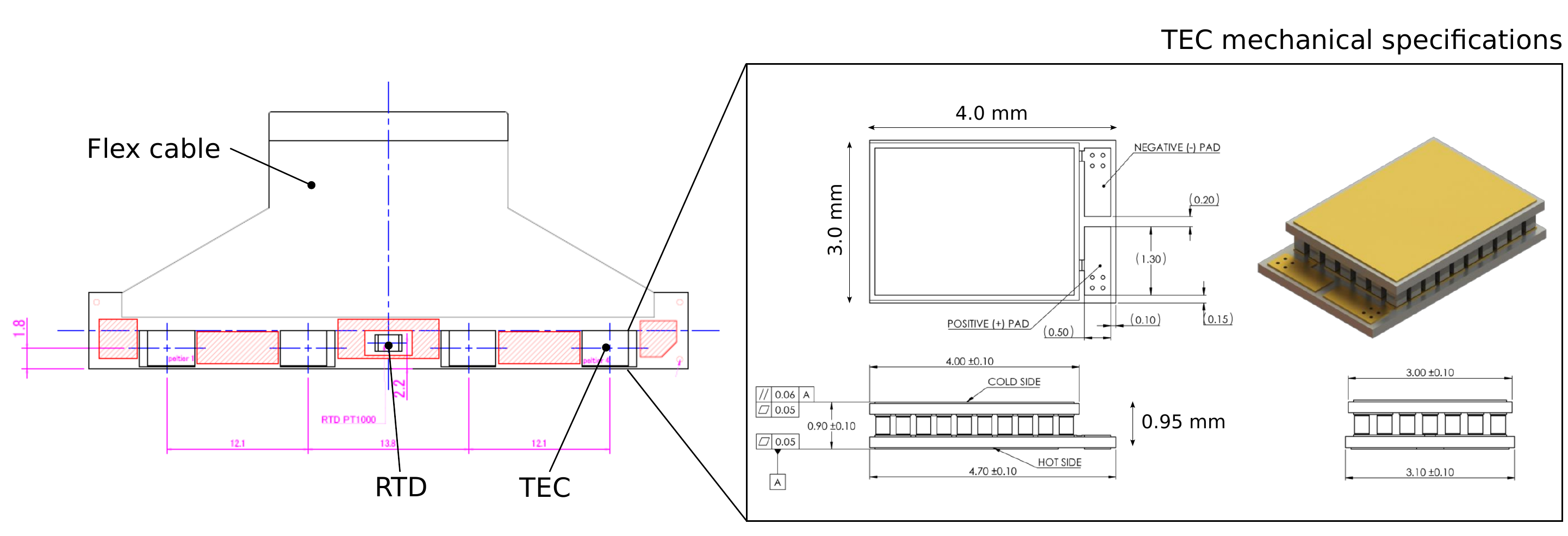}   \\
    \includegraphics[height=0.34\linewidth]{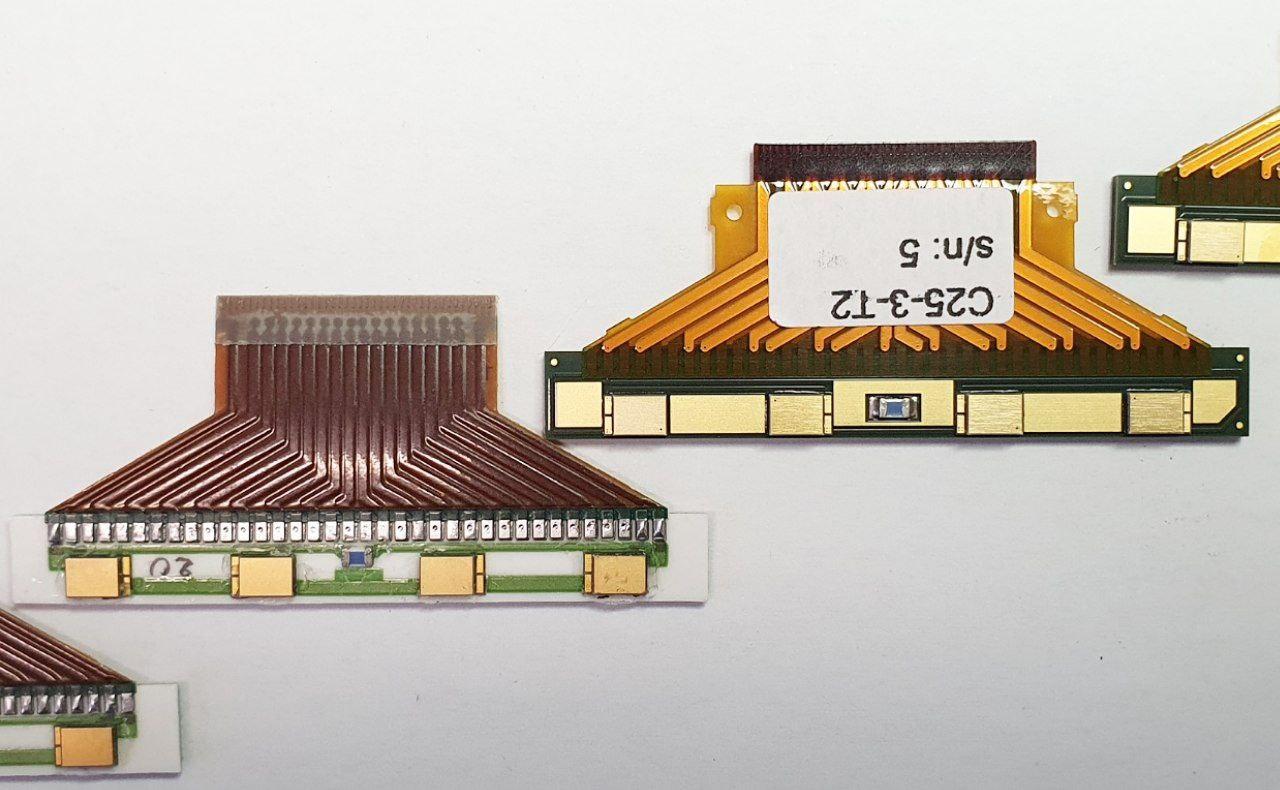}
    \includegraphics[height=0.34\linewidth]{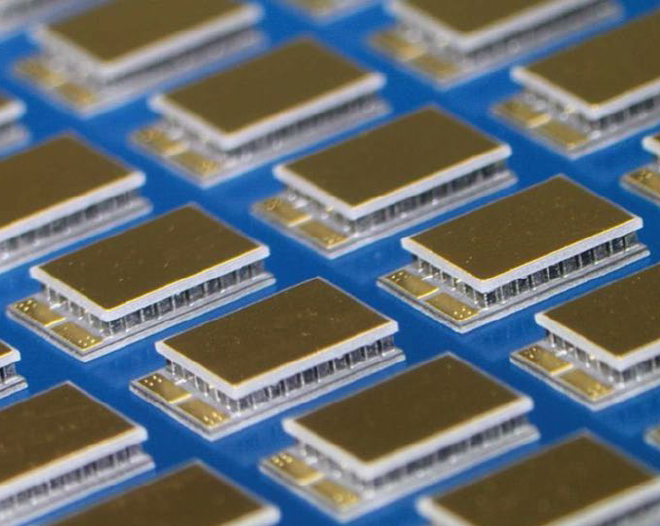}
    \caption{Top: Technical drawing of the back side of a BTL SiPM array where the location of the temperature sensor (RTD) and of the four TECs is visible, including dimensions of the thermo-electric coolers. Bottom: picture of BTL SiPM arrays (back side) from different manufacturers with TECs and RTDs (left) and picture of a batch of TECs as received from the supplier.}
    \label{fig:tecs_on_sipms}
\end{figure}

A BTL SiPM array consists of a 16-channel linear array with a 3.2~mm pitch and a height of about 3.0~mm. The total width of the package is 51.5~mm and the height is about 6.5~mm to house the SiPMs and their wire bondings. A flex cable is soldered on the rear side of the package to connect the individual SiPMs to the electronics (as also bias voltage is provided). An RTD is soldered in the center of the package below the flex cable as shown in figure~\ref{fig:tecs_on_sipms} to monitor the operating temperature. 

Driven by the environmental challenges discussed in the previous section, a series of small thermo-electric coolers (TECs) based on Bismuth Sesquitelluride (Bi$_{2}$Te$_{3}$) have also been included as an integrated component of the SiPM package. The TECs are re-flow soldered at 220\degC to the rear side of the SiPM package on dedicated solder pads. The TECs specifications can be found on the manufacturer (Phononic) website \cite{Phononic_datasheet}.
In the presence of TECs the heat produced by the SiPM can flow only through the TEC surfaces (rather than through the entire substrate). It is therefore important to use a high conductivity thermal gap filler to maximize the heat flow between the TEC side coupled to the SiPM substrate and the side coupled to the copper housing. A THERMFLOW Phase-Change Thermal Interface Pad (T558) \cite{T558} with thickness of about 50-150~$\rm \mu m$ was found as a good compromise between high thermal conductivity and capability to absorb potential ruggedness or non-planarity of the TEC and copper housing surfaces.

Besides cost and integration implications another key parameter to consider is the additional power required by the TECs to provide an additional temperature gradient of about 10\degC. This is because the total heat produced by a detector module ultimately needs to be extracted by the CO$_2$ cooling system which has a finite capacity \cite{CMS_MTD_TDR}. The reason why TECs represent a very attractive solution for integration with irradiated SiPMs is that the power required to cool down the SiPMs by additional 10\degC compared to the cooling plate is almost entirely compensated by the decrease of SiPM leakage current.

To achieve a first proof-of-principle of this power compensation mechanism, a set of four off-the-shelf TEC prototypes of $2\times 2$~mm$^2$ area were mounted on SiPM arrays, irradiated to $2\times10^{14}$~\neut, annealed for 80 minutes at 60\degC and then glued to a LYSO crystal array to gauge the actual performance in cooling down a BTL module operated at around 1.0~V.
The TEC hot sides were thermally coupled to a large heat sink (emulating the detector cooling plate) and placed inside a climate chamber at a temperature of $-35$\degC.
The TEC current was then varied using a power supply to scan the temperature gradient between -35\degC and -50\degC while measuring both the SiPM leakage current and the injected TEC power.

\begin{figure}[!tbp]
    \centering
    \includegraphics[width=0.8\linewidth]{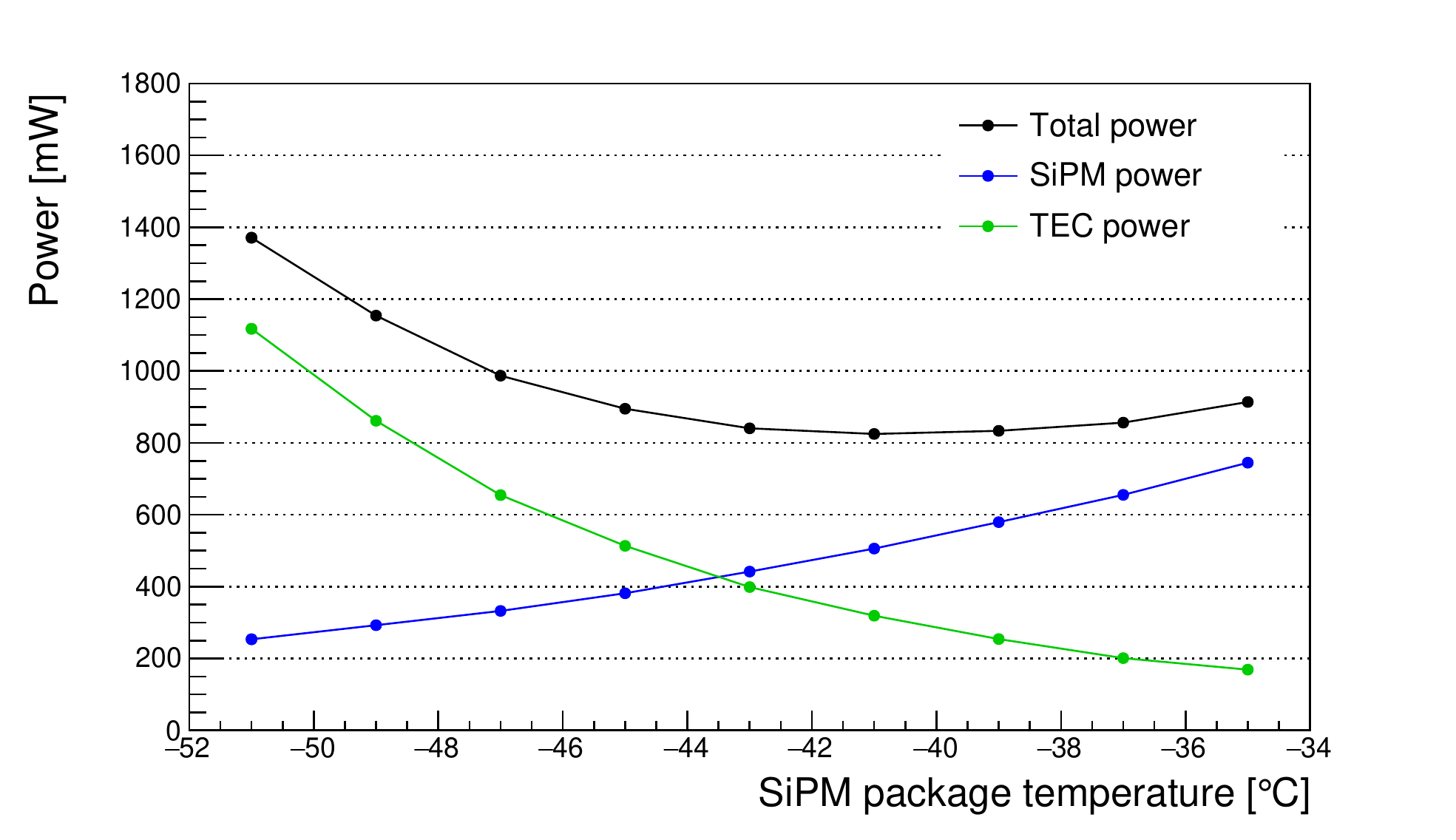}   \\
    \caption{Contribution of the SiPMs and the TECs to the total power consumption of a SiPM array with four $2\times2$~mm$^2$ TECs mounted on the package. The SiPM array was irradiated to $2\times10^{14}$~\neut~and annealed for 80 minutes at 60\degC. The measurement is performed by adjusting the bias voltage to maintain the SiPM over-voltage at 1~V for each temperature point.}
    \label{fig:tec_power_neutrality}
\end{figure}

As shown in figure~\ref{fig:tec_power_neutrality}, the total power budget of the SiPM+TECs system operating at a certain over-voltage is basically flat within 10\% for a SiPM operating temperature between $-35$\degC and $-45$\degC. In this sense the TECs enable the operation of SiPMs at a lower temperature and thus at a lower dark counts rate with a negligible impact on the total power budget of the detector. For a 10\degC gradient, the SiPM power (at fixed over-voltage) is lowered from 800~mW to 420~mW and the TEC power required is about 550~mW at $-35$\degC.

\noindent
Based on these first measurements on self wire bonded TEC prototypes ($2\times2$~mm$^2$ size) and a thermodynamic model of the system, described in section~\ref{sec:model_opt}, the number and the dimensions of the TECs were optimized. In particular, it was found that a total of four TECs with a $3\times4$~mm$^2$ surface area provides the best performance in terms of heat extraction within a fixed power budget. 

\noindent
Customized TECs of the desired dimensions have been produced by Phononic also using surface-mount device (SMD) technology with through via to ease integration on the SiPM package and improve robustness becoming a fully integrated part of the SiPM arrays for the BTL detector.
The voltage to the TECs is provided in series for the four elements on each package, given their high reliability discussed in section~\ref{sec:reliability}, and is routed through dedicated traces in the flex cable. As reported in figure~\ref{fig:tecs_on_sipms} the TEC thickness is limited to 0.95~mm thus remaining a compact solution with a small impact (about 3\%) on the detector acceptance. 

\section{TEC characterization and optimization}\label{sec:model_opt}

The performance of planar thermo-electric coolers, in terms of temperature gradient achieved for a certain electrical power, $P$, can be modeled by the following one-dimensional equations, assuming that most of the heat flow is orthogonal to the TEC largest surfaces with a temperature gradient, $\Delta T$ defined as the difference between the cold side, $T_c$, and the hot one, $T_h$:


\begingroup
\addtolength{\jot}{1em}
\begin{alignat}{1}
\setstretch{1.5}
& Q_c = \alpha I T_c - \frac{1}{2}R_{ac} I^2 - (K+K_{ev})\Delta T\label{eq:TECeqs}\\
& V                    = \alpha \Delta T + R_{ac} I\\
& P = VI
\end{alignat}
\endgroup

\noindent where $Q_c$ is the SiPM power (W), $\alpha$ the Seebeck coefficient (V/K), $R_{ac}$ the TEC resistance measured with an alternating current ($\Omega$), $I$ the TEC current (A) and $K$, the TEC conductance (W/K) which includes a correction for thermal leakage indicated as $K_{ev}$. For a given current flowing through the TEC, the three equations define the temperature gradient across the TEC, its voltage gradient, and its power consumption.

When the TECs are used during the detector operation to further cool down the SiPM temperature, the SiPM power represents a heat load on the cold side of the TECs which needs to be extracted. For a fixed bias voltage, the heat load is directly proportional to the SiPM dark current and thus depends on the effective over-voltage, operating temperature and annealing history. For the BTL detector at the end of operation $Q_c$ is estimated to be about 420~mW for the entire array (thus about 27~mW/SiPM) if operated at $-45^{\circ}$C, while it would be about 800~mW for the same operating voltage but a temperature of $-35^{\circ}$C.
Since the SiPM front side is in contact with the LYSO:Ce crystal through a 100$~\rm \mu m$ thick layer of silicone glue (RTV-3145) there will be thermal leakage (heat flow) from the crystal side as well as through the flex cable attached to the cold side of the package. From measurements obtained using a BTL detector module, the thermal leakage was estimated to be $K_{ev} = 63$~mW/\degC at $-35$\degC. 
The larger this thermal leakage the larger the power required by TEC to lower the SiPM temperature.


The effective Seebeck coefficient, thermal conductivity and electrical resistance of the TECs have been measured using the Harman's method \cite{Harman,Roh2016} with an injected current of 5~mA.
%
%
%
The measured values of $R_{ac}$, $\alpha$ and $K$ are reported in Table~\ref{tab:tec_pars} for both $2\times2$~mm$^2$ and $3\times4$~mm$^2$ TECs.
\begin{table}[!tbp]
\centering
\caption{TEC parameters used for modeling and simulation of $2\times2$~mm$^2$ and $3\times4$~mm$^2$ TECs measured with Harman's method with 5-10~mA.}
\vspace{0.2cm}
\begin{tabular}{c|c|c|c|c}
\hline
                &  \multicolumn{2}{c}{$3\times4$~mm$^2$ TEC} & \multicolumn{2}{c}{$2\times2$~mm$^2$ TEC}\\
                \hline
    Parameter   &  $T=25$\degC           & $T=-35$\degC &  $T=25$\degC           & $T=-35$\degC \\ 
    \hline \hline
    $R_{ac}$ [$\Omega$]   & 3.73 $\pm$ 0.07  & 2.58 $\pm$ 0.05 & 2.00 $\pm$ 0.04 & 1.40 $\pm$ 0.03\\ 
    $\alpha$ [mV/K]       & 14.7 $\pm$ 0.3   & 12.3 $\pm$ 0.2  & 7.0 $\pm$ 0.1  & 5.9 $\pm$ 0.1\\ 
    $K$      [mW/K]       & 22.2 $\pm$ 0.4   & 23.3 $\pm$ 0.5  & 10.9 $\pm$ 0.2  & 11.2 $\pm$ 0.2\\ 
    \hline
\end{tabular} 
\label{tab:tec_pars}
\end{table}
Both the resistance, $R_{ac}$, and the Seebeck's coefficient depend on temperature and so does the TEC coefficient of performance (COP) defined as the amount of heat exchanged ($\dot{Q}\equiv Q_c + K_{ev}\Delta T$) divided by the amount of supplied electrical power ($P=VI$).
For four TECs connected in series, as on the back of a SiPM package, the equivalent value of the coefficients is four times the one indicated in Table~\ref{tab:tec_pars}.
We can thus evaluate the expected power required to achieve a certain temperature gradient and the corresponding coefficient of performance, COP, for a generic number of TECs, $N_{tec}$, connected in series using the following equations:

\begingroup
\addtolength{\jot}{1em}
\begin{alignat}{1}\label{eq:1Dmodel}
\setstretch{1.5}
& \dot{Q} = N_{tec}\alpha(\bar{T}) I T_c - \frac{1}{2} N_{tec}R_{ac}(\bar{T}) I^2 - N_{tec} k\Delta T\\
& COP (\bar{T}) = \frac{\dot{Q}}{P}
\end{alignat}
\endgroup

\noindent
where $\bar{T} = (T_h+T_c)/2$ is the average temperature at which the TEC performance is evaluated.


%

\begin{figure}[!tbp]
    \centering
        \includegraphics[width=0.495\linewidth]{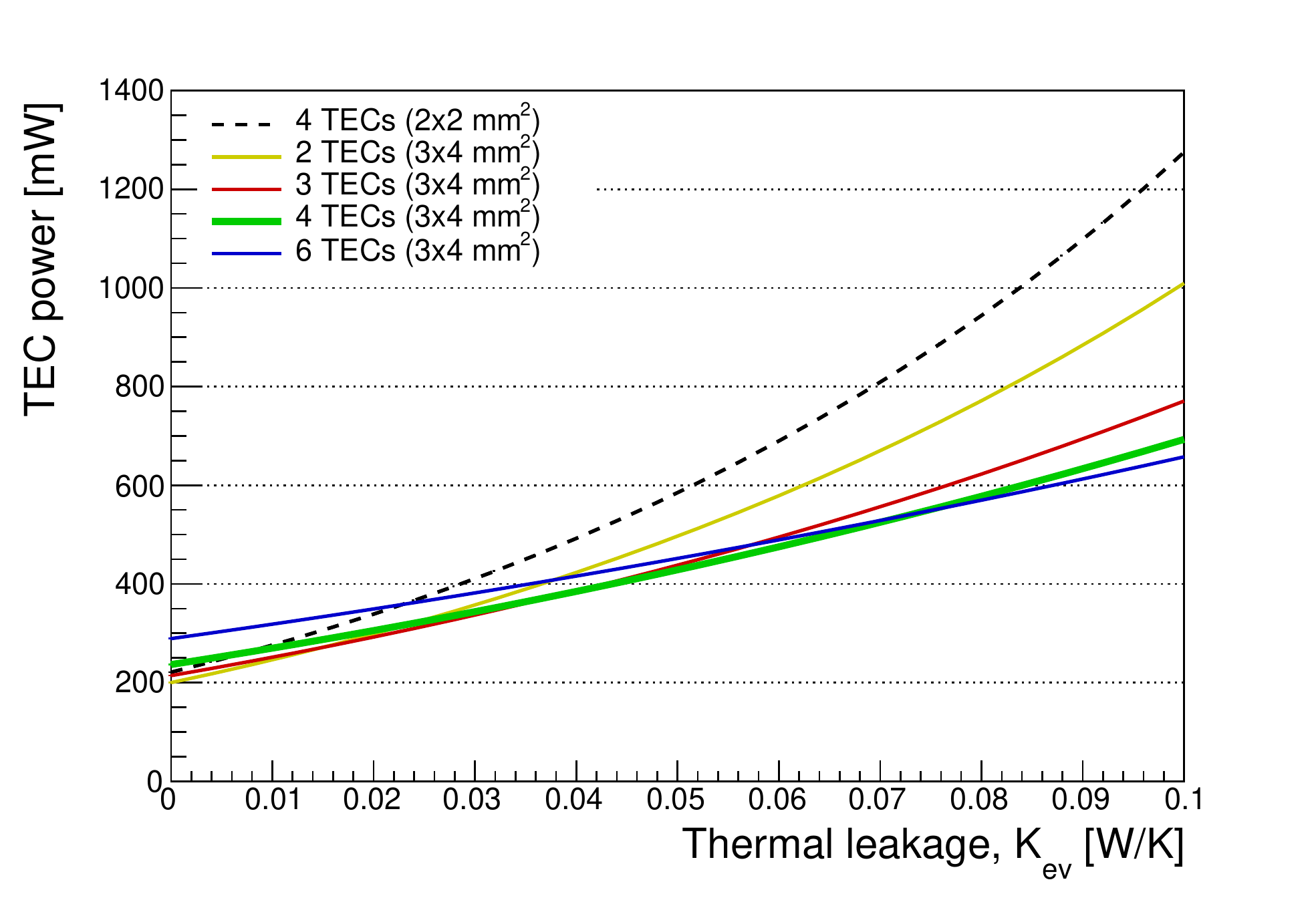}
        \includegraphics[width=0.495\linewidth]{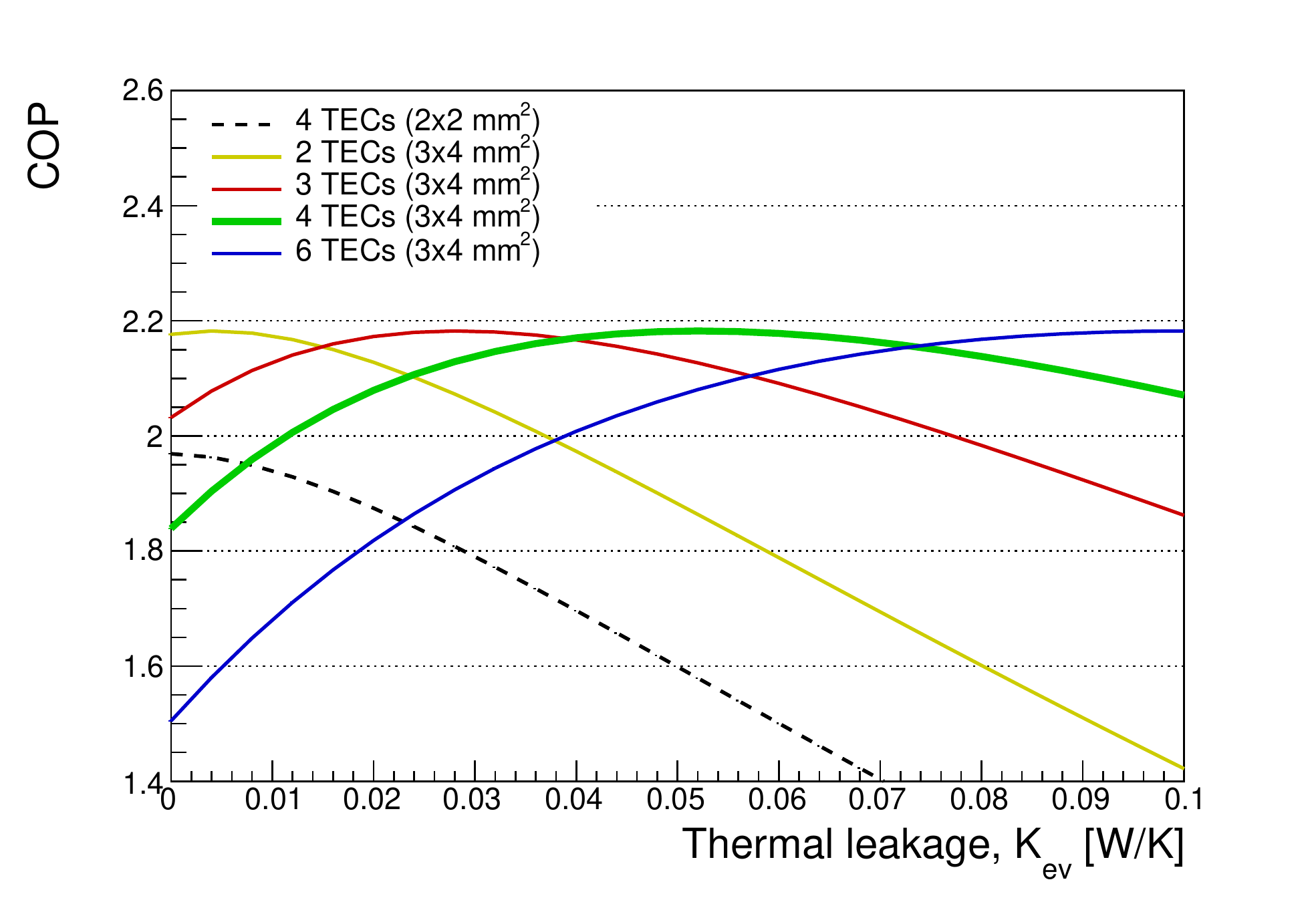}\\
        \includegraphics[width=0.495\linewidth]{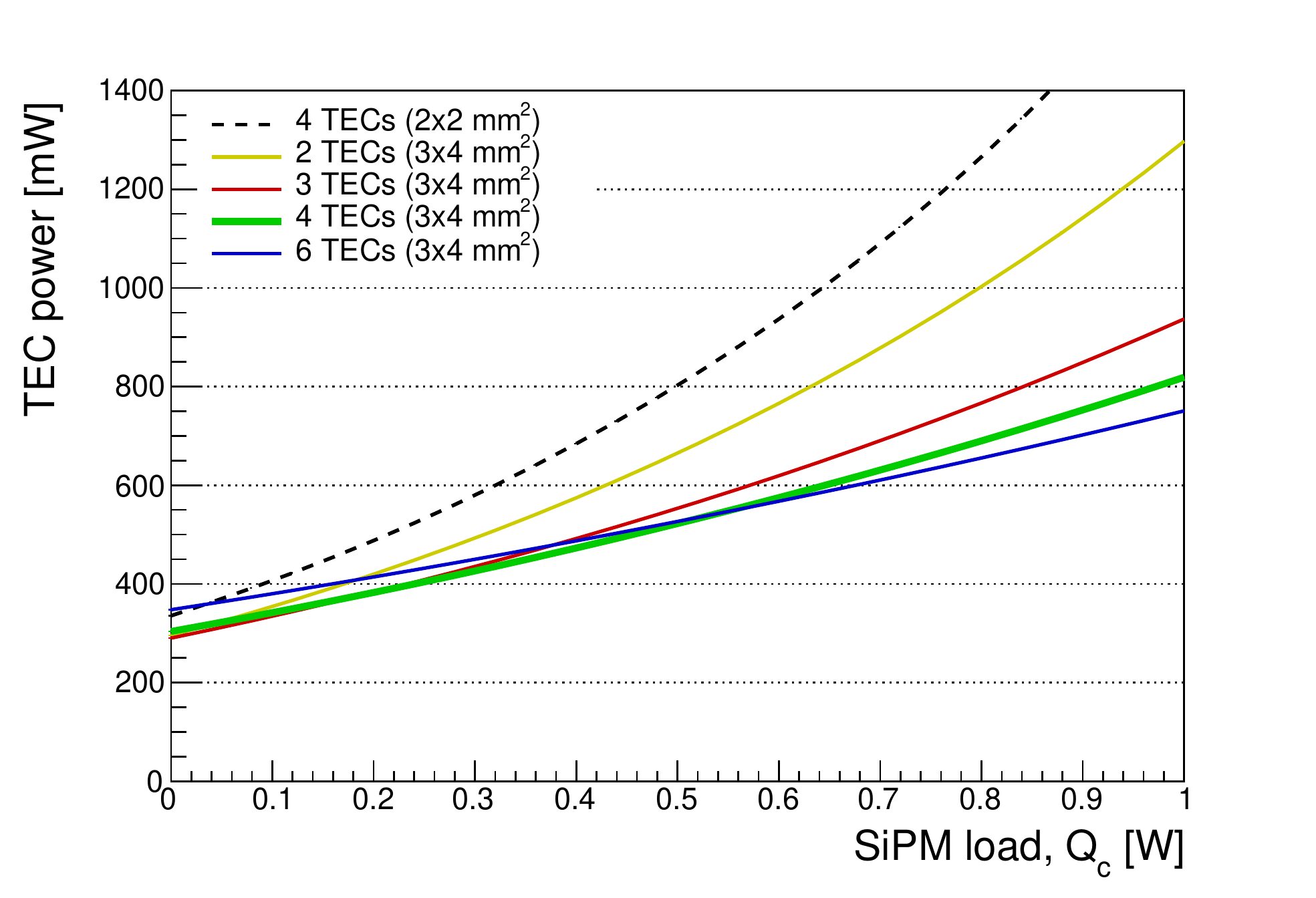}
        \includegraphics[width=0.495\linewidth]{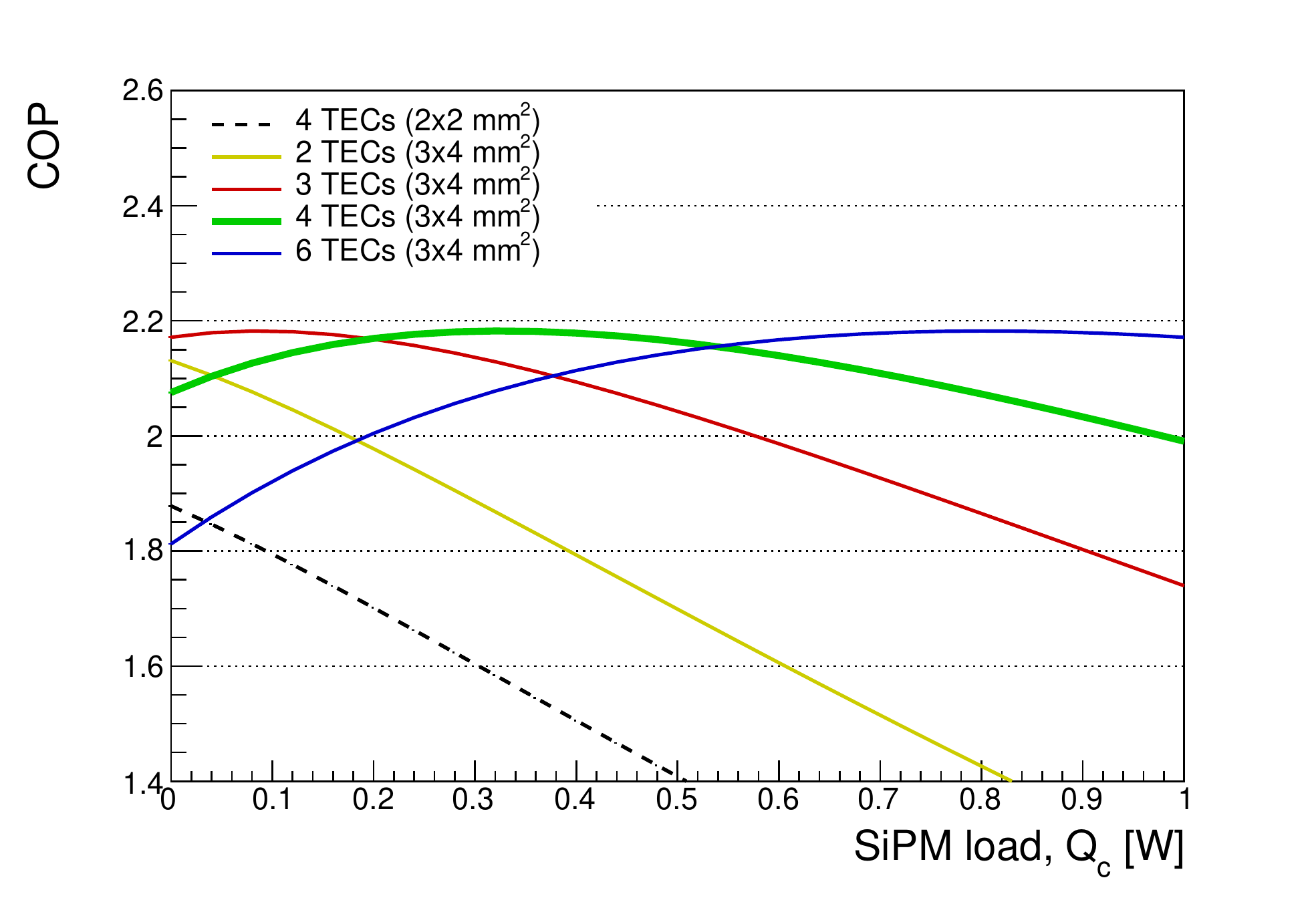}\\
    \caption{Top: Required power and COP to obtain a $\Delta T =-10$\degC at $-35$\degC as a function of the thermal leakage for different TEC configurations and a nominal SiPM load of 420~mW. Bottom: Required power and COP to obtain a $\Delta T=-10$\degC at $-35$\degC as a function of the SiPM load for different TEC configurations and a thermal leakage of 63~mW/\degC.}
    \label{fig:tec_optimization}
\end{figure}

Using the measured parameters for $2\times2$~mm$^2$ and $3\times4$~mm$^2$ TECs in Table~\ref{tab:tec_pars} and Eq.~\ref{eq:1Dmodel} we compared the expected TEC power required to achieve a 10\degC gradient and the corresponding COP for a few different configurations of TECs as a function of the SiPM load $Q_c$ and thermal leakage $K_{ev}$. The results are reported in figure~\ref{fig:tec_optimization} showing that TECs with a larger surface are more efficient in the case of large thermal leakage ($K_{ev}$) and large SiPM load ($Q_c$). Conversely, in the absence of thermal leakage and for small SiPM load smaller SiPMs ($2\times2$~mm$^2$) are more efficient.

\noindent
This was the major driver to develop custom $3\times 4$~mm$^2$ large TECs rather than using off-the-shelf products with $2\times 2$~mm$^2$ area.
Furthermore we concluded that there is no advantage in including more than four TECs on the package since as shown in figure~\ref{fig:tec_optimization}, for the given values of $K_{ev}$ and $Q_c$ of the BTL modules and a desired $\Delta T = -10^{\circ}$C, the 4-TECs configuration has the maximum COP.

\begin{figure}[!tbp]
    \centering
    \includegraphics[width=0.495\linewidth]{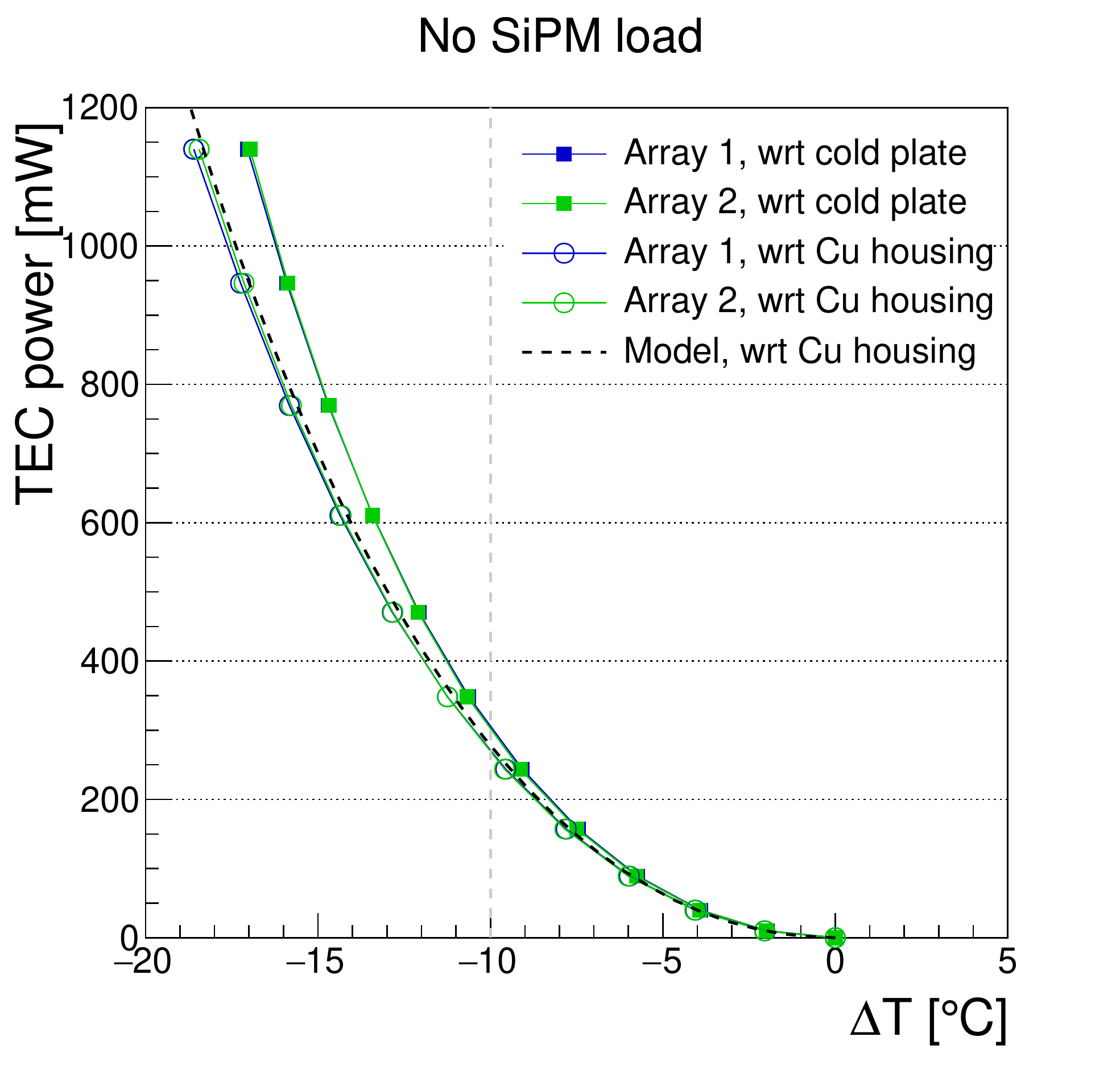}  
    \includegraphics[width=0.495\linewidth]{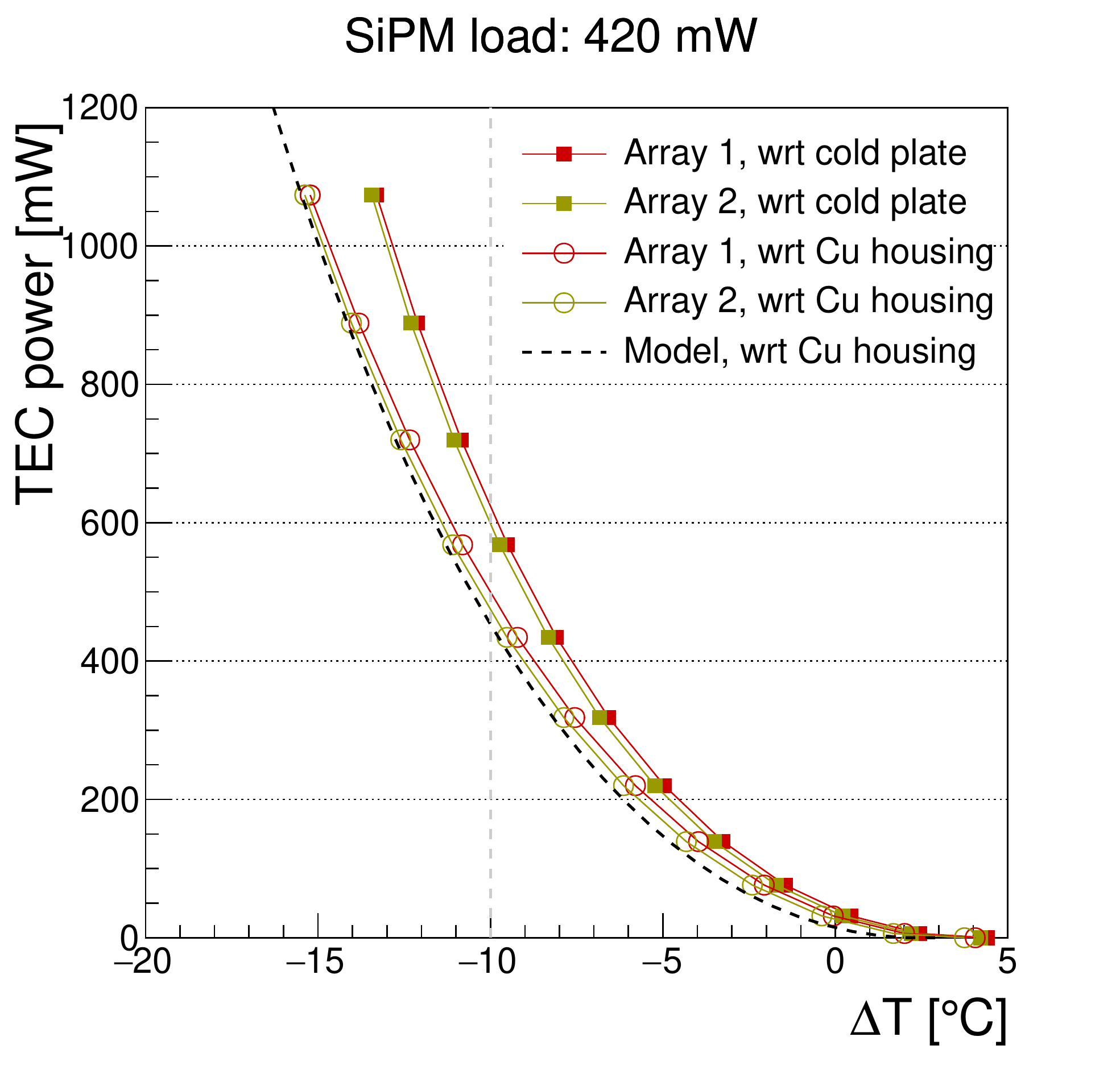}  \\
    \caption{Power consumption of four $3\times 4$~mm$^2$ TECs connected in series on two SiPM arrays attached to a LYSO module (array 1 and array 2) as a function of the temperature gradient achieved with respect to the copper housing (open dots) and to the cold plate (full dots). Left figure shows the power required when no SiPM load is present and the right figure when there is a SiPM load of 420~mW. The temperature of the copper housing was set to $-25$\degC.}
\label{fig:tec_performance}
\end{figure}

The performance of four $3\times4$~mm$^2$ TECs mounted on a SiPM array was characterized on a detector module inserted in a copper housing and attached to a cold plate.
The power required to achieve a certain temperature gradient with respect to both the copper housing and cold plate is shown in figure~\ref{fig:tec_performance}. The temperature of the copper housing was set at $-25$\degC and the measurement was performed both with and without SiPM load.
The results confirm a good agreement between the measurements and the 1D model. As expected the presence of SiPM load requires additional power to achieve the same temperature gradient.
It can also be noted that due to imperfect thermal coupling between the copper housing and the cold plate, an additional 100~mW are required to offset the additional 2\degC between the copper housing and the cold plate.

\begin{figure}[!tbp]
    \centering
    \includegraphics[width=0.495\linewidth]{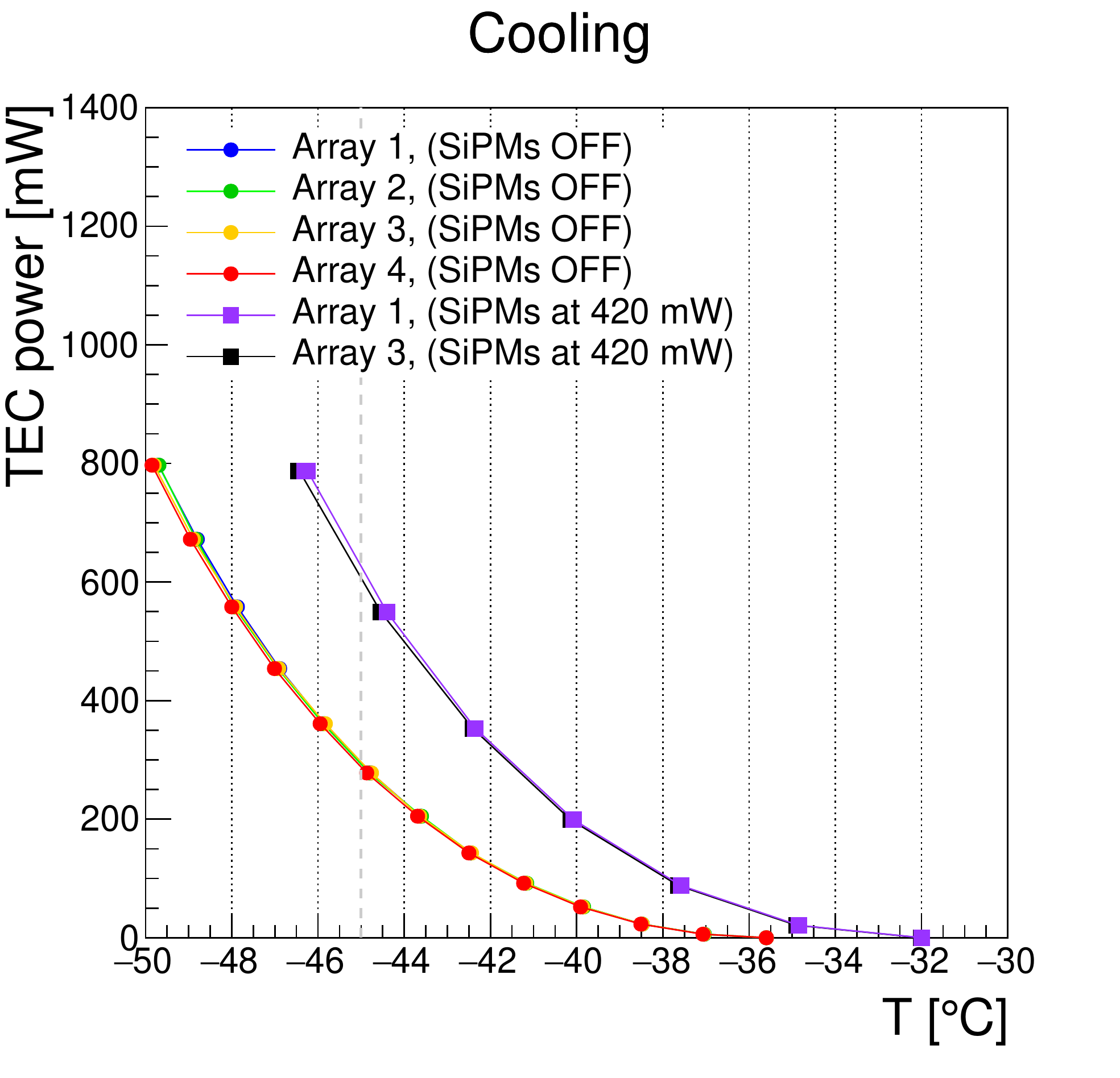}  
    \includegraphics[width=0.495\linewidth]{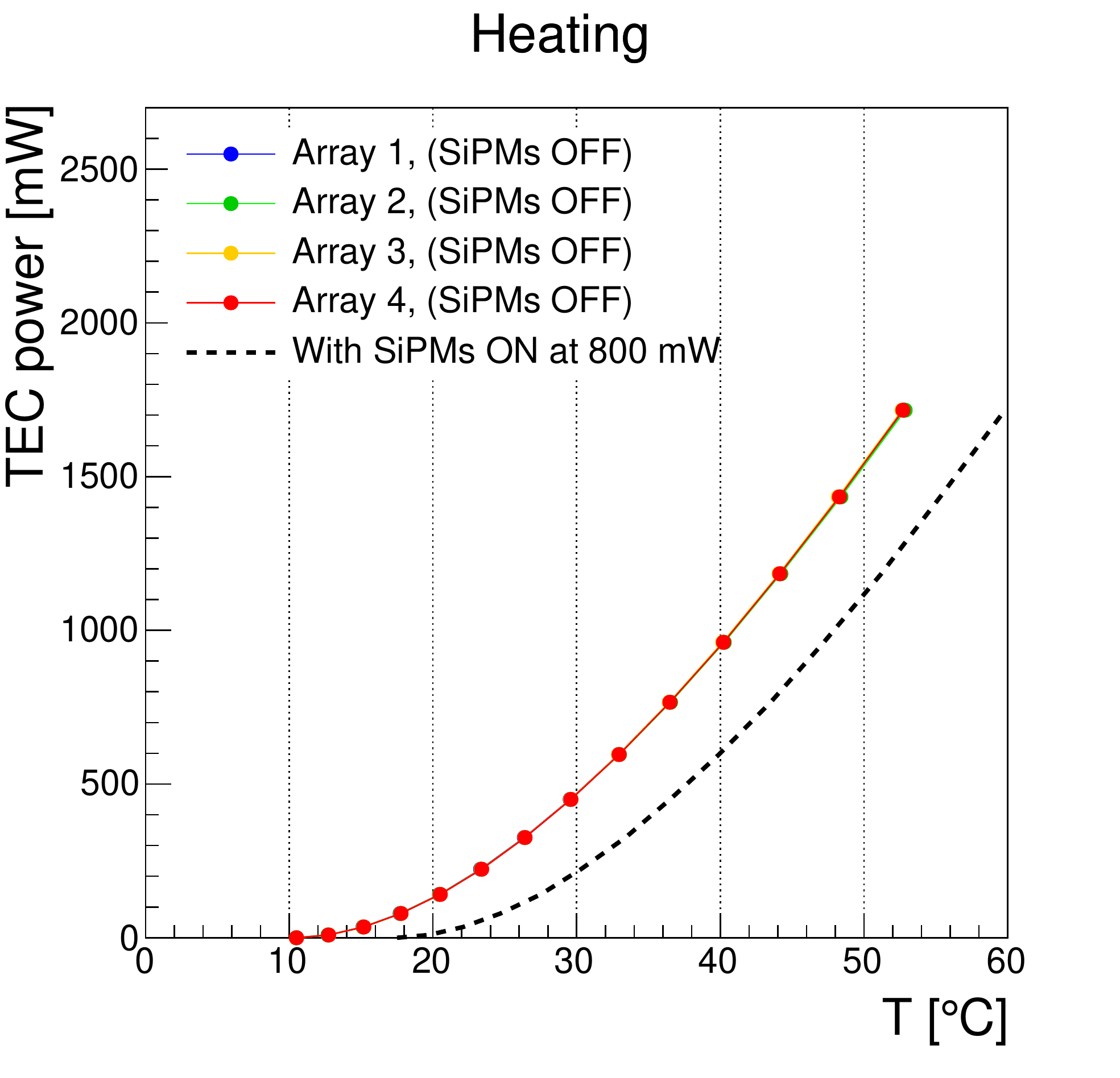}  
    \caption{TEC power consumption per SiPM array measured on the four arrays inside a detector module (array 1,2,3,4) as a function of the SiPM temperature achieved. Left figure shows the power required for the cooling configuration with the cold plate at $-35$\degC (with and without SiPM load). Right figure shows the power required for the heating configuration with the cold plate at $10$\degC (with and without SiPM load).}
\label{fig:tec_DM_performance}
\end{figure}

With the same procedure the power required to achieve the target SiPM temperature for the cooling and heating scenarios of the BTL detector was estimated and is reported in figure~\ref{fig:tec_DM_performance}. The performance of all four SiPM arrays inside the detector module was monitored.
For the cooling scenario, the cold plate was set to $-35$\degC and a SiPM temperature of $-45$\degC was achieved under a 420~mW SiPM load by injecting about 600~mW of TEC power biased with default polarity.
For the heating scenario the cold plate temperature was set to $10$\degC and a SiPM temperature of $60$\degC was achieved when operating the SiPM at 800~mW and the TECs at 1700~mW with reversed polarity.
%


\section{Simulation of TEC performance in BTL modules}\label{sec:tec_in_btl}

\begin{figure}[!htpb]
    \input{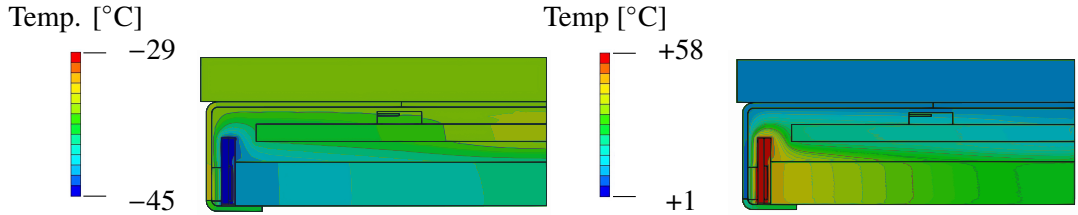}
	\caption{Simulation of the temperature inside a BTL detector module obtained using ANSYS (side view, see figure~\ref{fig:btl_module_section} for description of the components). The left plot shows the temperature map inside a module during operation with a cold plate at $-35$\degC and SiPMs at $-45$\degC, the right plot shows the temperature map during annealing periods, i.e. with the cold plate at 10\degC and the SiPMs at 60\degC.}
	\label{fig:tec_fem_simulation}
\end{figure}

A dedicated 3D finite element method (FEM) model was derived for the TEC elements and verified against the simple 1D model from Eq.~\ref{eq:1Dmodel} focused on replicating the TEC experimental setup and better understanding the heat flux inside a detector module. For this purpose, the ANSYS simulation software was used to fit the experimental temperature and power data and evaluate the thermal conductivity, heat transfer coefficient and thermal resistances of the assembly used in an adaptive single-objective optimization.
The simulated temperature within a BTL detector module mounted on the cooling plate are shown in figure~\ref{fig:tec_fem_simulation}. 
The left panel illustrates the temperature profile of a module cross-section operating in cold with a SiPM heat injection of 405~mW in the SiPMs and 460~mW per 4 TECs achieving a final temperature of $-45$\degC in the SiPM package. 
The right panel shows the temperature distribution inside a detector module under the heating scenario, where the highest temperature of 58\degC is reached at the SiPMs and the lowest temperature is at the opposite side of the TECs, next to the module's copper housing. 
The FEM also provided insights for optimization of the overall module design, in particular by optimizing the main thermal interfaces where most of the heat re-flow occurs and avoiding thermal short circuits.

\begin{figure}[!htpb]
    \input{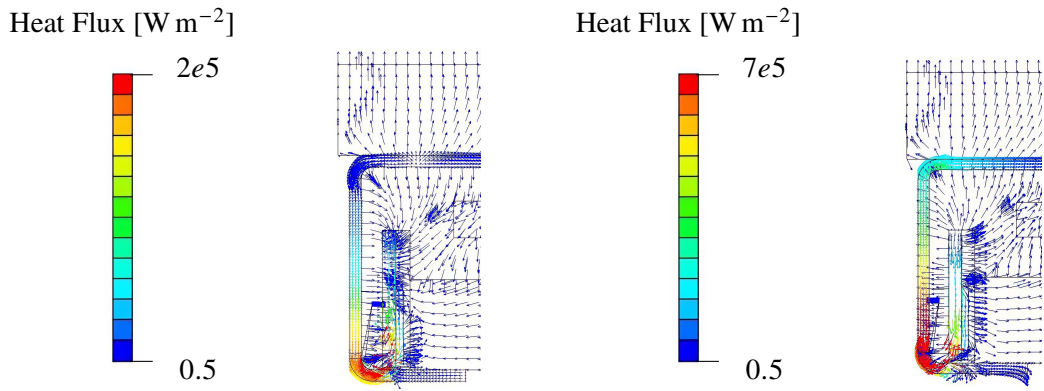}
	\caption{Simulation of the heat flow inside a BTL detector module obtained using ANSYS (side view, see figure~\ref{fig:btl_module_section} for description of the components). The left plot shows the vector heat flow inside a module during operation with a cold plate at $-35$\degC and SiPMs at $-45$\degC, the right plot shows the vector heat flow during annealing periods, i.e. with the cold plate at 10\degC and the SiPMs at 60\degC.}
	\label{fig:btl_heat_flows}
\end{figure}

Figure~\ref{fig:btl_heat_flows} shows the direction of the heat in the cold and annealing operation. The picture clearly shows the heat leaks through the air surrounding the SiPMs and towards the crystal.
The thermal contact between the TEC and the copper housing should also be maximized while a thermal insulation of the flange on the bottom side of the copper housing supporting the SiPM array should be made to avoid a thermal short-circuit.
%
%


\section{Uniformity, radiation tolerance and longevity tests}\label{sec:reliability}

More than 5000 custom TECs have been produced by Phononic. A fraction of these were mounted on dedicated PCBs for initial testing while others were mounted on BTL SiPM arrays by two different SiPM manufacturers (\vendorF~and \vendorH).
Measurements of the TEC parameters (Seebeck's coefficient and electric resistance) confirmed an excellent uniformity of the devices (less than 1\% standard deviation). Consistently, the uniformity of the temperature gradient achieved by different TECs biased with the same voltage is within $\pm 1$\degC up to a $\rm \Delta T$ of more than 25\degC.

Additional crucial requirements set on the TEC by the harsh operation environment of the BTL detector are excellent radiation tolerance, high reliability and longevity since the detector will not be accessible for repair during the entire decade of operation of the HL-LHC.

\begin{figure}[!tbp]
    \centering
    \includegraphics[width=0.495\linewidth]{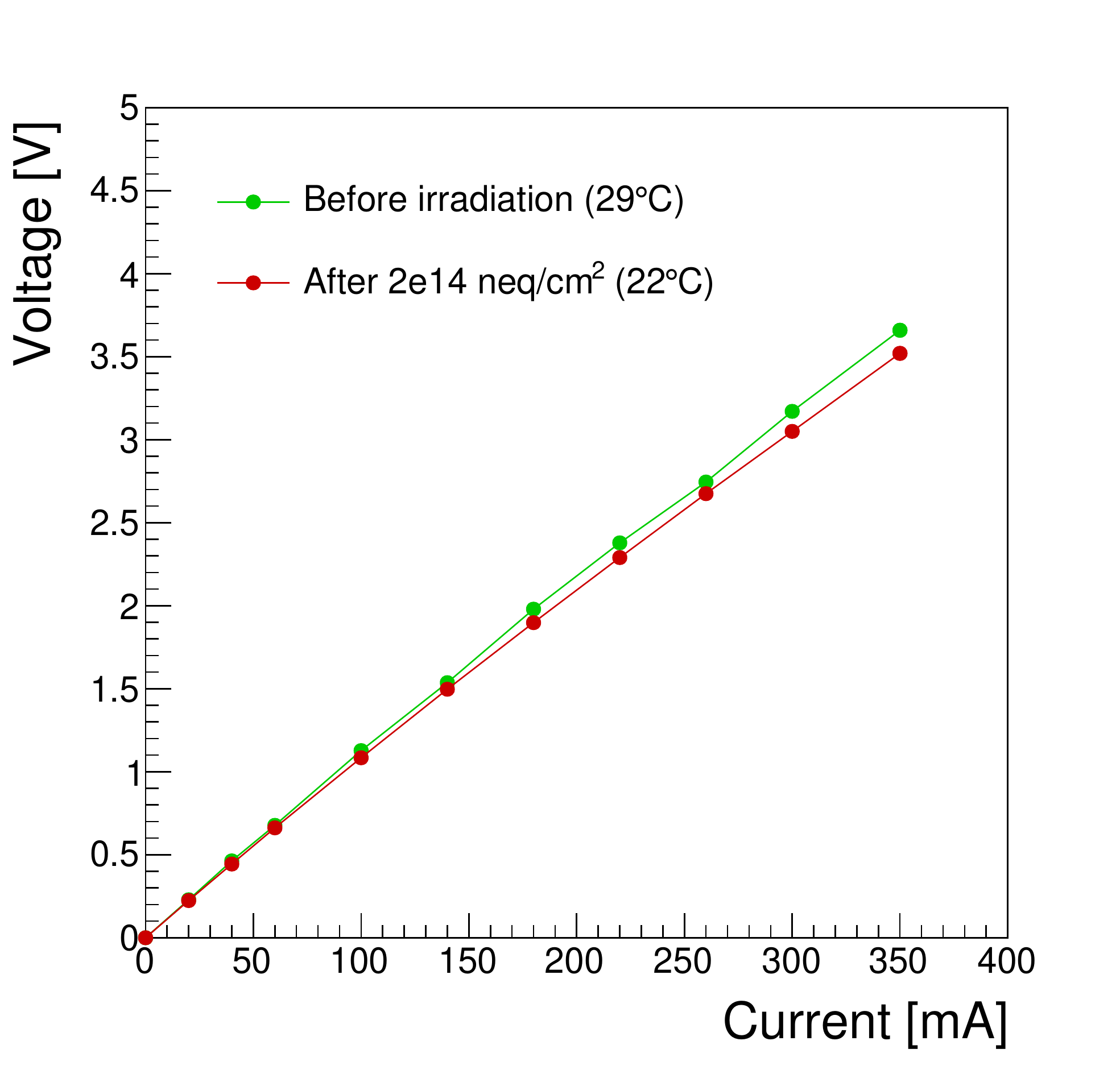} 
    \includegraphics[width=0.495\linewidth]{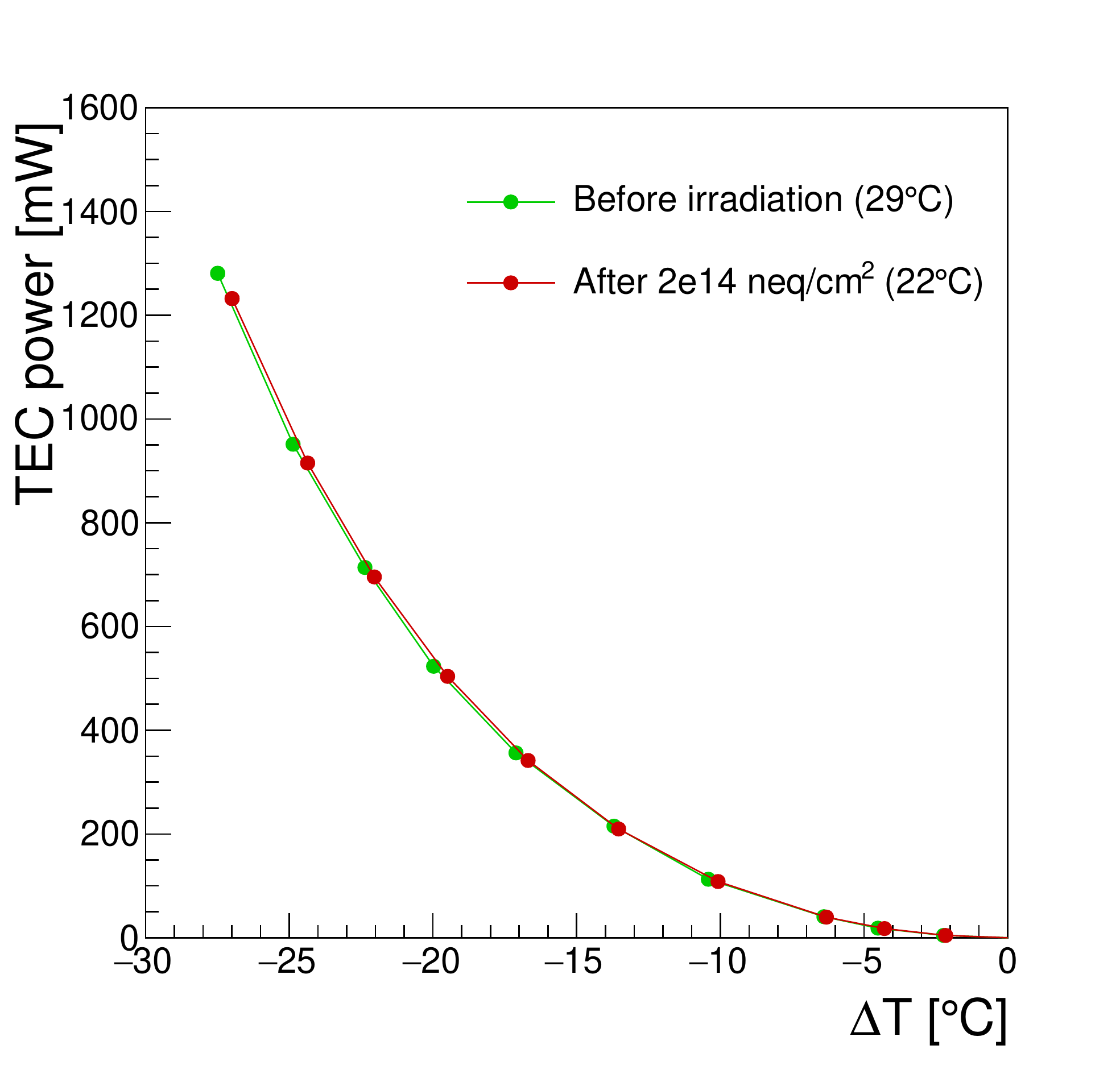}  \\
    \caption{Measured voltage as a function of current (left panel) for four TECs connected in series on a SiPM package before (green) and after irradiation to $2\times10^{14}$~\neut~(red) and corresponding TECs power as a function of the temperature gradient across the hot and cold surfaces (right panel).}
    \label{fig:tec_radtest}
\end{figure}

Since the TECs will be mounted on the SiPM package, they will be exposed to the same amount of radiation as the SiPMs themselves during HL-LHC running. To test the radiation hardness, measurements of the prototype TEC parameters and performance were thus done before and after irradiation to $2\times10^{14}$~\neut. Figure \ref{fig:tec_radtest} shows the results for the TEC resistance and power consumption as a function of temperature change. There is no sign of any radiation damage to the TECs. The small difference between results before and after irradiation is due to the fact that the measurements were done at slightly different temperatures. This excellent radiation tolerance is consistent with the high doping of such semiconductors (Bi$_2$Te$_3$) which are thus only negligibly affected by changes to the doping concentration that could be induced by radiation.

To determine the long-term reliability of the TECs, about $10^4$ thermal cycles have been performed. In particular, 1000 cycles were performed while keeping the TEC SiPM side at a temperature of $-43$\degC in a climate chamber and biasing the TEC to cycle between $-55$\degC and $-22$\degC and 10000 cycles were performed with the TEC SiPM side at 50\degC and biasing the TEC to cycle between 25\degC and 75\degC. The cycles have been performed at a rate of 3 minutes/cycle on 48 TECs simultaneously. No failures and no changes in the TEC parameters (e.g. the electrical resistance) were observed suggesting that the quality of the solder joints for instance is not affected by thermal stress. 
The absence of failures observed is consistent with the failure in time rate (FIT) claimed by the manufacturer of less than one TEC over $10^9$ hours of operation.

The TECs mechanical specifications tolerate a compression force of 25~N and 10~N for shear stress. The copper housing structures have thus been designed such that the force exerted by the housing on each TECs is about 10~N thus with a factor 2.5 safety compared to the TEC tolerance while ensuring a sufficient thermal contact. 
For our special application, in addition to the standard testing performed in accordance to Telcordia GR-468 platform for optoelectronic TECs 20000 cycles under a compression of 24.5~N were performed by Phononic on 11 TECs at the maximum power of 0.95~A and $\rm T_{hot}>75$\degC.
Possible damage to the TECs due to shipping, handling and installation in the modules can be easily quality controlled using a hand held LCR meter to check the internal resistance (with SDEV of 1\% by the manufacturer) and will be part of the detector quality control procedure during construction.

\section{Conclusions}

To address the unprecedented challenged posed by the operation of SiPMs in the harsh radiation environment ($2\times 10^{14}$~\neut~cumulative fluence) of the BTL detector, the integration of small thermo-electric coolers (TECs) into the SiPM package has been investigated.
Customized TECs produced by Phononic have been tested on a BTL detector module consisting of irradiated SiPM arrays glued to LYSO:Ce crystal arrays and inserted into a dedicated copper housing for heat dissipation. The experimental results have shown that the TECs are capable of providing additional reduction of the SiPM operating temperature from $-35$\degC to $-45$\degC without requiring additional power budget for the detector module. This results in a beneficial reduction of the dark count rate during operation by a factor of about two. 
Similarly, TECs have been proven capable to increase the SiPM operating temperature up to about 60\degC boosting the thermal annealing of the radiation damage effects and further recovering the performance of irradiated SiPMs. The combination of the deeper cooling and higher annealing temperature enabled by the TECs leads to an overall reduction of the dark count rate by about one order of magnitude.
The possibility to perform such thermal cycles without detrimental effects on the BTL detector components has been assessed through dedicated studies.
A set of full detector modules, including the wrapped LYSO crystal arrays glued to the SiPM arrays with TECs, was tested throughout 100 thermal cycles between $-50$\degC and $+60$\degC inside a climate chamber and no sign of mechanical instability nor effect on the light output and time performance of the modules was observed.

Following the positive results of dedicated radiation tolerance and longevity tests performed on the TECs, the integration of such devices on the SiPM package has been adopted as the baseline for the BTL detector. Several hundreds of prototype SiPM arrays have been produced by different SiPM manufacturers with the inclusion of thermo-electric coolers as an integrated component of the SiPM package and are being tested in view of the beginning of the BTL detector construction that will require a production of about 22000 SiPM arrays and 88000 TECs.

\acknowledgments

The addition of TECs on the SiPM arrays is a proposal of A.Heering, A.Karneyeu, Y.Musienko, and M.Wayne, followed by qualification and prototyping studies. The feasibility of integration in the MTD design is a combined effort, coordinated by M.Wayne and P.Meridiani, with contributions from W.Lustermann and K.Stachon for the biasing scheme, J.C.Rasteiro Da Silva and J.Varela for readout and control, A.Bornheim, G.Reales, I.Schmidt for mechanics and thermal integration aspects. The model of SiPM response under irradiation and annealing is a contribution of Y.Musienko, A.Ledovskoy, and A.Benaglia, based on specific irradiation campaigns carried out by A.Heering, Y.Musienko and T.Anderson, B.Cox, A.Ledovskoy, C.E.~Perez~Lara and S.White. Prototyping and validation on full size modules is a contribution of A.Benaglia, F.De Guio, A.Ghezzi, M.Lucchini, M.Malberti, S.Palluotto, T.Tabarelli de Fatis, M.Benettoni, R.Carlin, R.Rossin, M.Tosi, R.Paramatti and F.Santanastasio.
The authors would like to acknowledge the Quality Assurance and Reliability Testing (QART) laboratory of the EP-DT group at CERN for the support and equipment used during the tests as well as the personnel of the JSI facility in Ljubljana where SiPMs and TECs were irradiated. 
We are grateful to M.Moll who kindly loaned the climate chamber used in some of the tests.
We also thank Phononic for the development of custom thermo-electric coolers and for openly sharing specifications and results of in-house tests performed on the devices.

\bibliography{mybibfile}

\end{document}